\documentclass[reprint,floatfix,amsmath,amssymb,superscriptaddress
]{revtex4-2}

\usepackage{graphicx}
\usepackage{bm}
\usepackage{epstopdf}
\usepackage{xr}
\usepackage{float}
\usepackage{lineno}
\usepackage{siunitx}
\DeclareSIUnit\Oersted{Oe}
\DeclareSIUnit\electronvolt{eV}
\DeclareSIUnit\emu{emu}
\usepackage{xcolor}
\usepackage{mathtools, cuted}
\begin{document}


%\title{Enhancing the spin Hall effect in paramagnetic Co$_3$Sn$_2$S$_2$-based shandite thin films via Fermi-level tuning near the Dirac point}
\title{Intercorrelated anomalous Hall and spin Hall effect in kagome-lattice Co$_3$Sn$_2$S$_2$-based shandite films}

% author & affiliation
\author{Yong-Chang Lau}
\email{yongchang.lau@iphy.ac.cn}
\affiliation{Institute for Materials Research, Tohoku University, Sendai 980-8577, Japan}
\affiliation{Center for Spintronics Research Network, Tohoku University, Sendai 980-8577, Japan}
\affiliation{Institute of Physics, Chinese Academy of Sciences, Beijing 100190, China}
\author{Junya Ikeda}
\affiliation{Institute for Materials Research, Tohoku University, Sendai 980-8577, Japan}
\author{Kohei Fujiwara}
\affiliation{Institute for Materials Research, Tohoku University, Sendai 980-8577, Japan}
\author{Akihiro Ozawa}
\affiliation{Institute for Materials Research, Tohoku University, Sendai 980-8577, Japan}
\author{Takeshi Seki}
\email{takeshi.seki@tohoku.ac.jp}
\affiliation{Institute for Materials Research, Tohoku University, Sendai 980-8577, Japan}
\affiliation{Center for Spintronics Research Network, Tohoku University, Sendai 980-8577, Japan}
\author{Kentaro Nomura}
\affiliation{Institute for Materials Research, Tohoku University, Sendai 980-8577, Japan}
\affiliation{Center for Spintronics Research Network, Tohoku University, Sendai 980-8577, Japan}
\author{Atsushi Tsukazaki}
\email{tsukazaki@tohoku.ac.jp}
\affiliation{Institute for Materials Research, Tohoku University, Sendai 980-8577, Japan}
\affiliation{Center for Spintronics Research Network, Tohoku University, Sendai 980-8577, Japan}
\affiliation{Center for Science and Innovation in Spintronics, Core Research Cluster, Tohoku University, Sendai 980-8577, Japan}
\author{Koki Takanashi}
\affiliation{Institute for Materials Research, Tohoku University, Sendai 980-8577, Japan}
\affiliation{Center for Spintronics Research Network, Tohoku University, Sendai 980-8577, Japan}
\affiliation{Center for Science and Innovation in Spintronics, Core Research Cluster, Tohoku University, Sendai 980-8577, Japan}

\date{\today}

% abstract
\begin{abstract}
Magnetic Weyl semimetals (mWSMs) is characterized by linearly dispersive bands with chiral Weyl node pairs associated with broken time reversal symmetry. %\cite{VafekRev2014,YanRev2017,RevModPhys_WSM,Nagaosa_NRM2020}.
One of the hallmarks of mWSMs is the emergence of large intrinsic anomalous Hall effect. %\cite{YangPRB_AHEinWSM,Liu2018}.
On heating the mWSM above its Curie temperature, the magnetism vanishes while exchange-split Weyl point pairs collapse into doubly-degenerated gapped Dirac states.
Here, we reveal the attractive potential of these Dirac nodes in paramagnetic state for efficient spin current generation at room temperature via the spin Hall effect. %\cite{RevModPhys_SHE}.
Ni and In are introduced to separately substitute Co and Sn in a prototypal mWSM Co$_3$Sn$_2$S$_2$ shandite film and tune the Fermi level. %\cite{Liu1282,Morali1286}
Composition dependence of spin Hall conductivity for paramagnetic shandite at room temperature resembles that of anomalous Hall conductivity for ferromagnetic shandite at low temperature; exhibiting peak-like dependence centering around the Ni-substituted Co$_2$Ni$_1$Sn$_2$S$_2$ and undoped Co$_3$Sn$_2$S$_2$ composition, respectively.
The peak shift is consistent with the redistribution of electrons' filling upon crossing the ferromagnetic-paramagnetic transition, suggesting intercorrelation between the two Hall effects.
Our findings
%extend the scope of mWSMs to spintronic applications, for temperatures beyond their Curie points and
highlight a novel strategy for the quest of spin Hall materials, guided by the abundant experimental anomalous Hall effect data of ferromagnets in the literature.

\end{abstract}

% PACS numbers

%\maketitle
\maketitle

%main
%\section{Introduction}
\clearpage
Non-trivial topology in the band structure of a solid can give rise to large Berry curvature \cite{Berry1984,RevModPhys_Berry} acting as an effective magnetic field in real space. This field can deflect the electrons in motion, leading to an intrinsic off-diagonal transport contribution that does not depend on the extrinsic electrons' scattering rate.
Typical examples are the anomalous Hall effect (AHE) \cite{RevModPhys_AHE} in ferromagnets and its spin counterpart the spin Hall effect (SHE) \cite{RevModPhys_SHE}. The latter often involves non-magnetic metals with strong spin-orbit coupling and allows generation of a transverse spin current capable of manipulating the magnetization of an adjacent nanomagnet. The resulting spin-orbit torques (SOT) \cite{Manchon_SOTreview} is promising for applications including non-volatile memory, magnetic logic, field sensing and neuromorphic computing. Finding material systems that exhibit high charge-to-spin conversion efficiency is a key to realize competitive spin-orbitronic devices with low power consumption.
%The evaluation of AHE for a conductive magnetic material is relatively trivial and applicable to slab-shaped samples of any size because the electric charge obeys the continuity equation.
%In contrast, the spin current is not conservative and is subjected to dephasing when flowing through a material, typically over a nanometric length scale known as the spin diffusion length $\lambda$.
%Integration into nanodevices with one length scale comparable to $\lambda$ is therefore a prerequisite for SHE quantification.

SHE has thus far only been investigated for a small subset of all the known materials. One primary challenge for probing the SHE is the non-conservative nature of the spin current, thus necessitates material integration into devices of comparable length scale (e.g., the thickness) with the commonly nanometric spin diffusion length $\lambda$\cite{Bass_2007}. In contrast, owing to its ease of evaluation and prolonged history, the AHE for many conducting magnetic materials are known and available in the literature.
%Despite of the similar parsing and scaling relationship between these two phenomena, they had long been treated separately until more recently, the interplay between the magnetic order and the spin current generation has attracted an upsurge of interest.
In view of the very similar origin and scaling relationship for AHE \cite{OnodaPRB2008} and SHE \cite{Moriya2022}, it is tempting to study the intercorrelation between them. Using the abundant AHE data as a facile indicator, one may unveil new design principles for materials with large SHE.
Different from the trivial electronic bands in conventional 3d ferromagnetic metals \cite{Omori_PRB_2019}, we focus on one of the topological bands in a magnetic Weyl semimetal (mWSM) \cite{VafekRev2014,YanRev2017,RevModPhys_WSM,Nagaosa_NRM2020} prototype cobalt shandite Co$_3$Sn$_2$S$_2$ (CSS) to reveal the intriguing correlation between the AHE and SHE.
%A previous study on conventional 3d ferromagnetic metals, however, found no simple relationship between AHE and SHE \cite{Omori_PRB_2019}, possibly due to the trivial multi-band transport contributions in metallic systems. Here, we focus on semimetallic cobalt shandite Co$_3$Sn$_2$S$_2$ (CSS) exhibiting simpler electronic band structure and giant AHE to reveal the intriguing correlation between the AHE and SHE.

%On the other hand, experimental studies that makes use of these topological features in the band structure for harnessing a transverse spin current via the SHE are scarce. Unlike the charge-based transport phenomena, spin current is not conservative and is subjected to dephasing upon traversing an interface or upon flowing across a material over a characteristic length scale known as the spin diffusion length. Primary challenges of the experimental observations include the necessity to integrate these new materials into thin film devices and

\begin{figure*}
\begin{center}
  \includegraphics[width=1.8\columnwidth]{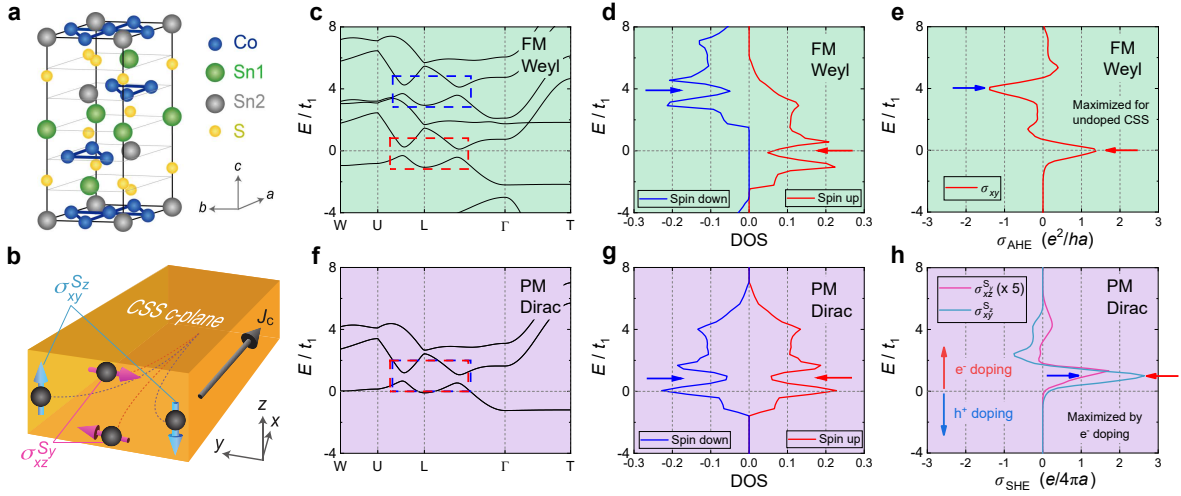}
  \caption{%\textbf{Concept of Fermi-level dependent Hall conductivities in Co$_{3}$Sn$_2$S$_2$.}
  (\textbf{a}) Unit cell of Co$_{3}$Sn$_2$S$_2$ with a Co-kagome lattice in the (ab) plane (hexagonal representation). (\textbf{b}) Passing a charge current $j_\textrm{c}$ within the kagome plane along $\textbf{x}$ generates, via the spin Hall effect, an orthogonal spin current of conductivity $\sigma_{xy}^{S_z}$ ($\sigma_{xz}^{S_y}$) that flows along $\textbf{y}$($\textbf{z}$) with polarization along $\textbf{z}$($\textbf{y}$). (\textbf{c-e}) Calculated electronic band structure (\textbf{c}), spin-resolved density of states (DOS) (\textbf{d}), and anomalous Hall conductivity $\sigma_{xy}$ (\textbf{e}) as a function of Fermi energy for ferromagnetic (FM) Co$_{3}$Sn$_2$S$_2$ in the magnetic Weyl semimetal state.
  %Near $E_\textrm{F}$, the presence of the Weyl points give raise to a local minimum in DOS and a peak in $\sigma_{xy}$.
  (\textbf{f-h}) Calculated electronic band structure (\textbf{f}), spin-resolved DOS (\textbf{g}), and two selected components $\sigma_{xy}^{S_z}$ and $\sigma_{xz}^{S_y}$ of the spin Hall conductivity tensor (\textbf{h})
   %for spin current flowing within the Kagome planes along $\textbf{y}$ (\textbf{i}) and for spin current flowing across the Kagome planes along $\textbf{z}$ (\textbf{j})
   for paramagnetic (PM) Co$_{3}$Sn$_2$S$_2$. $E/t_1 = 0$ represents the Fermi level $E_\textrm{F}$ for undoped Co$_{3}$Sn$_2$S$_2$ in the FM and PM states, respectively. Red and blue dashed boxes in (\textbf{c}) and (\textbf{f}) represent the position of the Weyl/Dirac points. The arrows in (\textbf{d}), (\textbf{e}), (\textbf{g}), and (\textbf{h}) indicate contributions of these Weyl/Dirac points in DOS and Hall conductivities.
  %Proper electron ($e^-$) doping in Co$_{3}$Sn$_2$S$_2$ is required to maximize the spin Hall conductivities.
  }
  \label{fig:schematic}
\end{center}
\end{figure*}

Figure~\ref{fig:schematic}(a) illustrates a rhombohedral structure of CSS (Space group No. 166; $R\bar{3}m$) which consists of alternate Co$_3$Sn/\allowbreak~SnS$_2$ planes, stacking along the $\textbf{c}$-axis in the hexagonal representation. Co atoms form a kagome lattice within the $\textbf{ab}$ plane and exhibit strong perpendicular magnetic anisotropy with Curie temperature ($T_\text{C}$) of $\sim$\SI{177}{\kelvin}. Recent spectroscopic studies \cite{Liu1282,Morali1286} have established ferromagnetic CSS (FM-CSS) as an exotic mWSM with pairs of Weyl points (WPs) near the Fermi level ($E_\text{F}$) connected by the chiral surface Fermi arcs. This leads to large summation of the Berry curvature in FM-CSS, exhibiting a record-high anomalous Hall angle exceeding 0.2 \cite{Liu2018,Wang2018}. The relatively low $T_\text{C}$ of CSS, however, hinders the prospect of exploiting its topological properties for many practical applications.
%At temperatures above $T_\text{C}$, the conjugated Weyl points of CSS merge to form doubly-degenerated Dirac points above $E_\text{F}$ (Fig.~\ref{fig:schematic}g). Paramagnetic pristine CSS may therefore be an unoptimized Dirac semimetal.
So far, very limited studies \cite{Li_CSS_oxydation} were devoted to explore the usefulness of paramagnetic CSS (PM-CSS) at room temperature. Here, we highlight the intercorrelation between the intrinsic AHE and SHE for CSS-based compounds, across the ferromagnetic-paramagnetic transition. We further demonstrate that Fermi-level tuning via isostructural substitutional alloying is an effective strategy allowing full exploitation of the topological features in the band structure of PM-CSS, for efficient spin current generation via SHE at room temperature.

To elucidate our concept, we consider a two-orbital effective tight-binding model of CSS \cite{OzawaJPSJ2019,Supplementary}. A longitudinal charge current $j_\textrm{c}$ is applied along $\textbf{x}$ in the $\textbf{ab}$ kagome plane, as illustrated in Fig.~\ref{fig:schematic}(b). The magnetization is along $\textbf{z}$ coinciding with the $\textbf{c}$-axis. We first focus on FM-CSS in the mWSM state, corresponding to Figs.~\ref{fig:schematic}(c)--\ref{fig:schematic}(e). Highlighted by the dashed red box in the simplified electronic band structure of half-metallic FM-CSS in Fig.~\ref{fig:schematic}(c), the two WPs in the vicinity of $E_\text{F}$ (represented by $E/t_1 = 0$), manifest themselves in the spin-resolved density of states (DOS) [Fig.~\ref{fig:schematic}(d)] as a minima of the majority spin-up DOS. The anomalous Hall conductivity $\sigma_{xy}$ (AHC) [Fig.~\ref{fig:schematic}(e)] exhibits a peak centering around the WPs at $E_\text{F}$. We highlight that at higher energy, a similar $\sigma_{xy}$ peak of opposite sign appears (blue arrow), which is attributed to the two conjugated WPs from the spin-down bands [Fig.~\ref{fig:schematic}(c); dashed blue box]. Using the lattice parameter of CSS $a=\SI{0.538}{\nano\meter}$ and recognizing $e^2/h$ is half of the conductance quantum, at $E_\text{F}$, we obtain $\sigma_{xy} \approx \SI{1.0E3}{\Omega^{-1}\centi\meter^{-1}}$, which agrees with the experiment and first-principles calculations \cite{Liu2018}.

\begin{figure*}
\begin{center}
  \includegraphics[width=1.6\columnwidth]{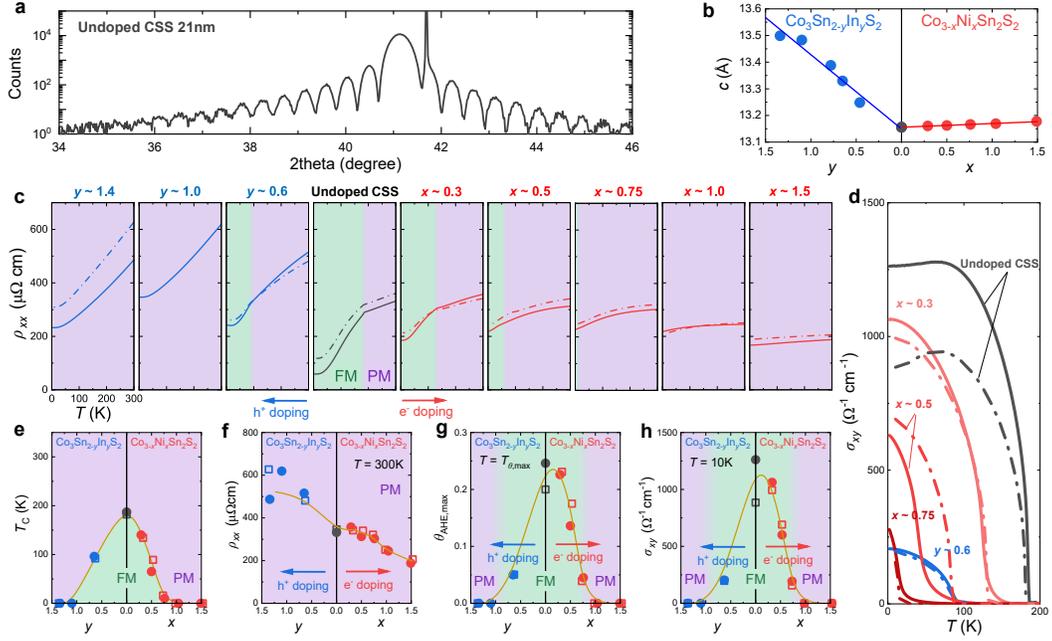}
  \caption{%\textbf{Structure and magneto-transport of Co$_{3}$Sn$_2$S$_2$-based single layers}
  (\textbf{a}) Typical x-ray diffraction spectrum for an undoped Co$_{3}$Sn$_2$S$_2$ shandite film showing clear Lau\'{e} fringes near the Co$_{3}$Sn$_2$S$_2$(0006) reflection. (\textbf{b}) Ni composition $x$ and In composition $y$ dependence of out-of-plane lattice parameter $c$ deduced from the peak position of the (0006) reflection. (\textbf{c, d}) Temperature $T$ dependence of the longitudinal resistivity $\rho_{xx}$ (\textbf{c}) and anomalous Hall conductivity $\sigma_{xy}$ (\textbf{d}) for shandite films of various $x$ and $y$. Solid and dashed lines represent data for the films before and after removing the thick SiO$_x$ capping, respectively. (\textbf{e-g}) $x$ and $y$ dependence of the Curie temperature $T_\textrm{C}$ (\textbf{e}), $\rho_{xx}$ at $T = \SI{300}{\kelvin}$ (\textbf{f}), the maximum anomalous Hall angle $\theta_\textrm{AHE,max} \equiv \sigma_{xy}\rho_{xx}$ extracted at $T = T_\mathrm{\theta,max}$ (\textbf{g}), and $\sigma_{xy}$ at \SI{10}{\kelvin} (\textbf{h}). Solid and open symbols denote data for shandite films before and after removing the thick SiO$_x$ capping, respectively. Dark yellow solid lines are drawn for guides to the eye.}
  \label{fig:singlelayer}
\end{center}
\end{figure*}

The calculated properties for PM-CSS where the exchange splitting is absent are shown in Figs.~\ref{fig:schematic}(f)--\ref{fig:schematic}(h). The conjugated WPs annihilate into doubly-degenerated Dirac points followed by gap opening due to spin-orbit coupling. [Fig.~\ref{fig:schematic}(f); dashed boxes] \cite{BelopolskiPRL2021,LiuDFPRB2021} The spin-resolved DOS of the two spins channels are equal, as shown in Fig.~\ref{fig:schematic}(g). $\sigma_{xy}$ vanishes because contributions from the two spins exactly cancel each other, as expected for a paramagnet. In contrast, the spin current is time reversal invariant, e.g., a spin-up spin current flowing along $\textbf{y}$ is equivalent to a spin-down spin current flowing along $-\textbf{y}$. The spin Berry curvature contributions from the two spin channels are additive and linked to the position of the gapped Dirac point. As depicted in Fig.~\ref{fig:schematic}(b), for $j_\textrm{c}$ along $\textbf{x}$, we define the spin Hall conductivity (SHC) $\sigma_{xi}^{S_j}$ ($i,j = x, y, z$) where the spin current is flowing along $\textbf{i}$ with polarization along $\textbf{j}$. Lateral SHC $\sigma_{xy}^{S_z}$ [Fig.~\ref{fig:schematic}(h); skyblue] of approximately twice-as-large to AHC is expected near the gapped Dirac point. Using the same $a$, we obtain $\sigma_{xy}^{S_z} \approx \SI{1.9E3}(\hbar/2e)\Omega^{-1}\text{cm}^{-1}$, which approaches that of a typical spin Hall metal Pt \cite{GuoPRL2008,Hoffmann}. In practice, for (0001)-textured CSS film in this work, it is more convenient to detect the spin current flowing along $\textbf{z}$ with polarization along $\textbf{y}$, i.e., $\sigma_{xz}^{S_y}$. Calculations suggest a peak near the gapped Dirac point for $\sigma_{xz}^{S_y}$ [Fig.~\ref{fig:schematic}(h); pink]. This maximum is however smaller than that of $\sigma_{xy}^{S_z}$ where charge and spin current are both flowing in the kagome plane, a feature resembles another kagome semimetal Fe$_3$Sn$_2$ with highly anisotropic AHC tensor \cite{YeNature2018}. The strong anisotropy reflects that the interplay between the conduction electron, kagome lattice, and spin-orbit coupling determines the Hall effects. We emphasize that SHC of undoped PM-CSS is unoptimized as $E_\text{F}$ is well below the gapped Dirac point. Within rigid band approximations, an electron doping of $\sim 1$ electron per formula unit is required to reposition $E_\text{F}$ close to the gapped Dirac point, for which a dramatic enhancement of SHC is expected.

We have grown textured undoped CSS, Ni--substituted Co$_{3-x}$Ni$_x$Sn$_2$S$_2$ (CNSS), and In--substituted Co$_3$Sn$_{2-y}$In$_y$S$_2$ (CSIS) films with thicknesses ranging from 11 to \SI{40}{\nano\meter} on Al$_2$O$_3$ (0001) substrates by co-sputtering as described previously and in Section S1 of the Supplemental Material \cite{Fujiwara_2019,Ikeda2021CommMater,Ikeda2021CommPhys,Supplementary}. $x$ and $y$ denote the composition of Ni and In, respectively. Upon replacing Co with Ni (Sn with In), the electron (hole) doping is expected to shift $E_\text{F}$ to higher (lower) energies. The crystal structure of CSS is maintained throughout because CSS, Ni$_3$Sn$_2$S$_2$ and Co$_3$In$_2$S$_2$ are isostructural compounds. Figure~\ref{fig:singlelayer}(a) shows typical x-ray diffraction (XRD) spectrum for an undoped CSS film exhibiting clear Lau\'{e} fringes around the CSS (0006) reflection, indicative of strong (0001) texture with sharp interfaces. In-plane XRD $\Phi$ scan of the CSS (11$\bar{2}$0) reflection, however, shows peaks that are \SI{30}{\degree} apart, reflecting the presence of in-plane twinned domains. The out-of-plane lattice parameter $c$ in Fig.~\ref{fig:singlelayer}(b) was extracted from the (0006) peak position and suggests systematic tuning of $x$ and $y$.

The magneto-transport as a function of temperature $T$ in Figs.~\ref{fig:singlelayer}(c) and \ref{fig:singlelayer}(d) provides another evidence of systematic tuning of film properties by alloying. Solid lines denote data for undoped CSS (black), CNSS (red) and CSIS (blue) films covered by a thick $\sim\SI{75}{\nano\meter}$ SiO$_x$ protective layer, whereas dashed lines represent the data for samples after removing the SiO$_x$ layer by Ar ion milling followed by the deposition of \SI{3}{\nano\meter} AlO$_x$ capping \cite{Supplementary}.
%We found slight degradation of the conductivity after the etching process while the general trend of the $T$ dependence remains unchanged.
%Kinks in the $T$ dependence of the longitudinal resistivity $\rho_{xx}$ (Fig.~\ref{fig:singlelayer}c) correspond to $T_\textrm{C}$ which varies systematically with $x$ and $y$.
The high quality of our undoped CSS film gives rise to a residual resistivity ratio, $\textrm{RRR} \equiv \rho_{xx}(T=\SI{300}{\kelvin})/\rho_{xx}(T=\SI{10}{\kelvin}) \sim 3$.
For doped films, the kink in $\rho_{xx}$ reflecting $T_\textrm{C}$ and the average resistance are systematically controlled by In and Ni contents.
Meanwhile, $\sigma_{xy}$ [Fig.~\ref{fig:singlelayer}(d)] is nearly a plateau for $T$ sufficiently far below $T_\textrm{C}$, suggesting the intrinsic nature of the AHE. $T_\textrm{C}$, $\rho_{xx}$ at $T = \SI{300}{\kelvin}$, the maximum anomalous Hall angle $\theta_\textrm{AHE,max} \equiv \sigma_{xy}\rho_{xx}$ at $T = T_\mathrm{\theta,max}$, and $\sigma_{xy}$ at \SI{10}{\kelvin} are summarized in Figs.~\ref{fig:singlelayer}(e)--\ref{fig:singlelayer}(h).
%$\rho_{xx}$ at $T = \SI{300}{\kelvin}$ decreases systematically with increasing Ni content.
Referring to Stoner criterion, both electron and hole doping reduce $T_\textrm{C}$ because $E_\text{F}$ of undoped PM-CSS falls on a local maxima of the DOS [Fig.~\ref{fig:schematic}(g)]. Similarly, $\sigma_{xy}$ at \SI{10}{\kelvin} and $\theta_\textrm{AHE,max}$ are maximized for undoped FM-CSS and fall rapidly with increasing $x$ and $y$. These observations confirm the correlation between the large intrinsic $\sigma_{xy}$ and the $E_\text{F}$ positioning relative to the magnetic WPs.

\begin{figure*}
\begin{center}
  \includegraphics[width=1.5\columnwidth]{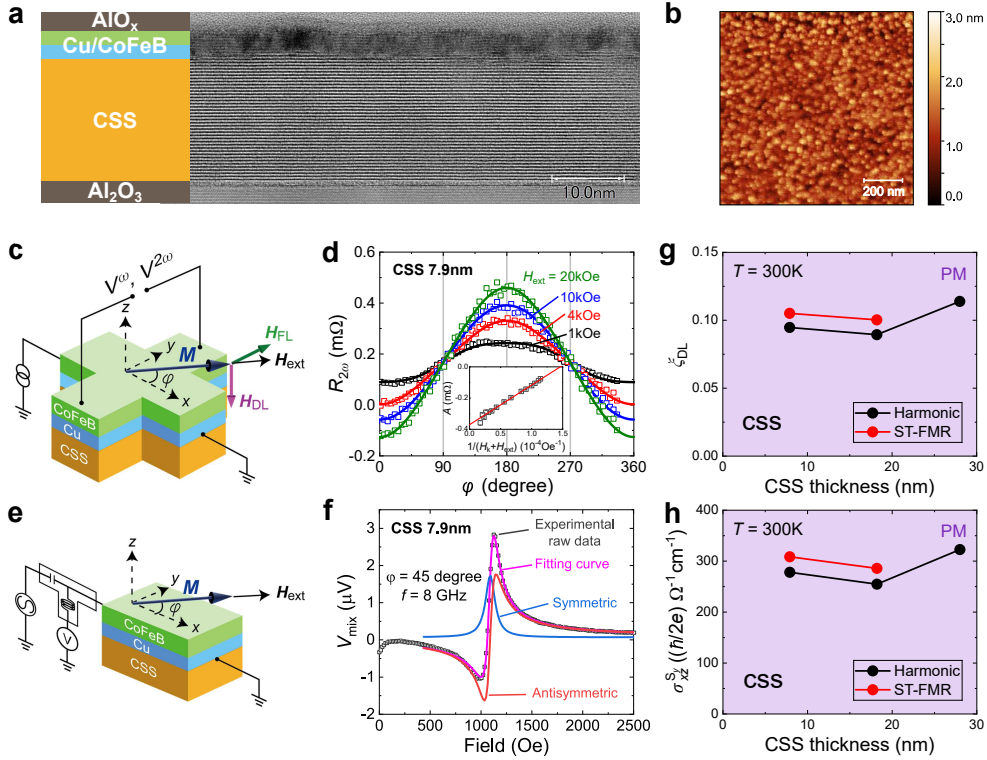}
  \caption{
  %\textbf{Structural characterization and spin-orbit torque quantification for undoped Co$_{3}$Sn$_2$S$_2$($t$)/Cu/CoFeB trilayers}
  (\textbf{a}) A schematic of the stack structure and high-resolution cross sectional transmission electron microscopy image for a trilayer with Co$_{3}$Sn$_2$S$_2$ (CSS) thickness $t = \SI{18.5}{\nano\meter}$. (\textbf{b}) Atomic force microscopy image for a trilayer with $t = \SI{7.9}{\nano\meter}$, showing a low root mean square roughness of $\sim\SI{0.3}{\nano\meter}$. (\textbf{c-f}) Spin-orbit torque quantification of Co$_{3}$Sn$_2$S$_2$($7.9$)/Cu/CoFeB trilayers at $T = \SI{300}{\kelvin}$. Schematic illustration of the harmonic Hall measurement set-up (\textbf{c}) and azimuthal field angle $\varphi$ dependence of the second harmonic Hall resistance $R_{2\omega}$ (\textbf{d}) measured at various external fields $H_\textrm{ext}$. Inset of (\textbf{d}) plots $A$, the prefactor of the $\cos{\varphi}$ term of $R_{2\omega}$ against the inverse of the effective field $1/(H_\textrm{k}+H_\textrm{ext})$. Red line is the best linear fit to the data. Schematic illustration of the spin-torque ferromagnetic resonance (ST-FMR) measurement set-up (\textbf{e}) and a typical FMR spectrum (\textbf{f}) measured at \SI{8}{\giga\hertz} with an applied field along $\varphi=\SI{45}{\degree}$. The fit and decomposition of the spectrum is based on the sum of a symmetric and an antisymmetric Lorentzian \cite{Supplementary}. (\textbf{g-h}) $t$ dependence of the damping-like spin Hall efficiency $\xi_\textrm{DL}$ (\textbf{g}) and spin Hall conductivity $\sigma_{xz}^{S_y}$ (\textbf{h}).}
  \label{fig:thickness}
\end{center}
\end{figure*}

We next fabricate undoped CSS($t$)/Cu(1.8)/CoFeB(2) (thicknesses in nanometer) trilayers [See Supplemental Material \cite{Supplementary}] for investigating the charge-to-spin conversion at $T=\SI{300}{\kelvin}$, i.e., when CSS is paramagnetic. $t$ denotes the thickness of the CSS layer. Figure~\ref{fig:thickness}(a) shows a typical high-resolution cross sectional transmission electron microscopy (HR-TEM) image of CSS(18.5)/Cu(1.8)/CoFeB(2) trilayer where the layered structure of high-quality CSS is clearly visible. The average grain size of CSS ($> \SI{50}{\nano\meter}$) is significantly larger than that of Cu/CoFeB ($\sim \SI{10}{\nano\meter}$). Energy dispersive x-ray spectroscopy (EDX) mapping confirms that all layers of the heterostructure are continuous with limited interdiffusion [See Fig.~S1 in Supplemental Material \cite{Supplementary}]. Typical atomic force microscopy (AFM) micrograph [Fig.~\ref{fig:thickness}(b)] of the trilayer for $t=\SI{7.9}{\nano\meter}$ reveals flat surface morphology with a low root mean square (r.m.s.) surface roughness of $\sim\SI{0.3}{\nano\meter}$, which is a prerequisite for spintronic device integration in the future.

On passing a charge current along $\textbf{x}$, CSS generates a spin current flowing along $\textbf{z}$ that traverses the Cu spacer and exerts SOTs on the CoFeB with in-plane magnetization. Limited by the thin film geometry, only $\sigma_{xz}^{S_i}; i = x,y,z$ are accessible. The thin Cu spacer with long $\lambda$ physically separates CSS and CoFeB, thus avoiding local enrichment of Co at the interface which may alter the properties of CSS. The observation of $\sim \SI{1}{\percent}$ current-in-plane giant magnetoresistance for the trilayer at $T=\SI{50}{\kelvin}$ (i.e., when CSS is ferromagnetic) confirms the finite spin transparency across the Cu spacer [See Fig.~S2 in Supplemental Material \cite{Supplementary}]. We first employ the harmonic Hall technique \cite{KimNatMat2012_Harmonic,GarelloNatNano2013_SOT} to quantify the
%charge-to-spin conversion. A sinusoidal current excitation applied to a Hall bar device generates
damping-like and field-like spin-orbit effective fields ($H_\textrm{DL}$ and $H_\textrm{FL}$, respectively) acting on the CoFeB magnetization [Fig.~\ref{fig:thickness}(c)].
%giving rise to an out-of-phase second harmonic Hall resistance $R_{2\omega}$ . We apply an external field $H_\textrm{ext}$ rotating within (\textbf{xy}) plane and making an angle $\varphi$ with the current,
The dependence of second harmonic Hall resistance $R_{2\omega}$ on the external field $H_\textrm{ext}$ and its azimuthal angle $\varphi$ allows separation of the SOT contribution ($\propto \frac{1}{H}$) from the parasitic thermoelectric effects \cite{AvciPRB2014_thermoelectric,RoschewskyPRB2019,ChiSciAdv2020_BiSb}. Figure~\ref{fig:thickness}(d) plots the $\varphi$-dependence of $R_{2\omega}$ measured at various $H_\textrm{ext}$ for trilayer with $t=\SI{7.9}{\nano\meter}$ and a current density flowing in the CSS of $j_\textrm{CSS} \sim \SI{1.6E6}{\ampere\per\centi\meter\squared}$. $R_{2\omega}(\varphi)$ is dominated by the $\cos{\varphi}$ term, defined with a prefactor $A$. [See Section S2 of the Supplemental Material for the detailed analysis of the harmonic Hall measurement \cite{Supplementary}]
%The counter-intuitive enhancement of $R_{2\omega}$ with increasing $H_\textrm{ext}$ is attributed to a strong thermoelectric background, which highlights the necessity of performing the full $H_\textrm{ext}$ dependence analysis.
$H_\textrm{DL}$ is extracted by linear fitting $A$ against the inverse of the effective in-plane field $1/(H_\textrm{k}+H_\textrm{ext})$ [insets of Fig.~\ref{fig:thickness}(d)]. We found $H_\textrm{DL}/j_\textrm{CSS} = \SI{1.4E-6}{\Oersted\per\ampere\centi\meter\squared}$, corresponding to a DL spin Hall efficiency $\xi_\textrm{DL} = \frac{2e}{\hbar} \frac{H_\mathrm{DL} M_\mathrm{s} t_\mathrm{CoFeB}}{j_\mathrm{CSS}} \approx +0.10$ where $e$ is the elementary charge, $\hbar$ the Planck constant, $M_\mathrm{s} = \SI{1200}{\emu\per\centi\meter\cubed}$ the saturation magnetization, and $t_\mathrm{CoFeB} = \SI{2}{\nano\meter}$ the CoFeB thickness. $\xi_\textrm{DL}$ of undoped PM-CSS is of the same sign as that of Pt. Its absolute magnitude is larger than another mWSM prototype Co$_2$MnGa ($\xi_\textrm{DL} \sim -0.07$ in the ferromagnetic state) \cite{Tang_APL2021}.
% and is comparable to that of $\beta$-Ta.
With $\rho_{xx} \approx \SI{340}{\micro\ohm\centi\meter}$, we obtained $\sigma_{xz}^{S_y} = \xi_\textrm{DL}/\rho_{xx} \approx 300 (\hbar/2e)\Omega^{-1}\text{cm}^{-1}$.

%$R_{2\omega}$ is of the form \cite{AvciPRB2014_thermoelectric,ChiSciAdv2020_BiSb}:
%\begin{equation}
%\begin{split}
%\label{eq:R2w}
%R_{2\omega}(H_\mathrm{ext},\varphi)=&\left(R_\mathrm{AHE}\dfrac{H_\mathrm{DL}}{H_\mathrm{ext}+H_\mathrm{k}}+R_\mathrm{ONE}H_\mathrm{ext}+R_\mathrm{const}\right)\cos\varphi\\
%           & \ \ -2R_\mathrm{PHE}\dfrac{H_\mathrm{FL}+H_\mathrm{Oe}}{H_\mathrm{ext}}\cos2\varphi\cos\varphi\\
%            &=R_{2\omega,\cos\varphi}\cos\varphi+R_{2\omega,\cos2\varphi\cos\varphi}\cos2\varphi\cos\varphi\\
%\end{split}
%\end{equation}
%$R_{\rm{AHE}}$ is the anomalous Hall resistance, $R_{\rm{PHE}}$ the planar Hall resistance, $H_{\rm{DL}}$ the damping-like (DL) spin orbit effective field, $H_{\rm{FL}}$ the field-like (FL) spin orbit effective field, $H_{\rm{Oe}}$ the Oersted field, and $H_{\rm{k}}$ the anisotropy field. $R_\textrm{const}$ and $R_\textrm{ONE}H_\textrm{ext}$ represent thermoelectric contributions that are independent of $H_\textrm{ext}$ (e.g. the anomalous Nernst effect or the combination of spin Seebeck effect and inverse spin Hall effect) and linear in $H_\textrm{ext}$ (e.g. the ordinary Nernst effect), respectively.

For an independent verification, we performed spin-torque ferromagnetic resonance (ST-FMR) [set-up depicted in Fig.~\ref{fig:thickness}(e)] measurement \cite{LiuPRL2011} on microstripes fabricated on the same substrate. A representative FMR spectrum measured from the mixing voltage $V_\textrm{mix}$ while applying $H_\textrm{ext}$ along $\varphi = \SI{45}{\degree}$ is shown in Fig.~\ref{fig:thickness}(f). Data is fitted by the sum of a symmetric and an antisymmetric Lorentzian. More details of the analysis can be found in the Section S3 of the Supplemental Material \cite{Supplementary}. The emergence of an appreciable symmetric component (blue) confirms the generation of DL-SOT from the undoped PM-CSS at room temperature. Line-shape analysis taking into account current shunting in the Cu spacer yields $\xi_\textrm{DL} = +0.11$, in excellent agreement with the harmonic Hall results. The full $\varphi$ dependence of ST-FMR confirms the dominant role of spin current with polarization along $\textbf{y}$ [See Fig.~S4(a) in Supplemental Material \cite{Supplementary}]. Figures~\ref{fig:thickness}(g) and \ref{fig:thickness}(h) summarize the CSS thickness $t$-dependence of $\xi_\textrm{DL}$ and $\sigma_{xz}^{S_y}$, obtained from the two techniques. The almost constant trend for $\xi_\textrm{DL}$ and $\sigma_{xz}^{S_y}$ against $t$ is consistent with bulk-like SHE and $\lambda$ much shorter than $t=\SI{7.9}{\nano\meter}$.
%governs the spin current generation in undoped PM-CSS.

\begin{figure}
\begin{center}
  \includegraphics[width=0.95\columnwidth]{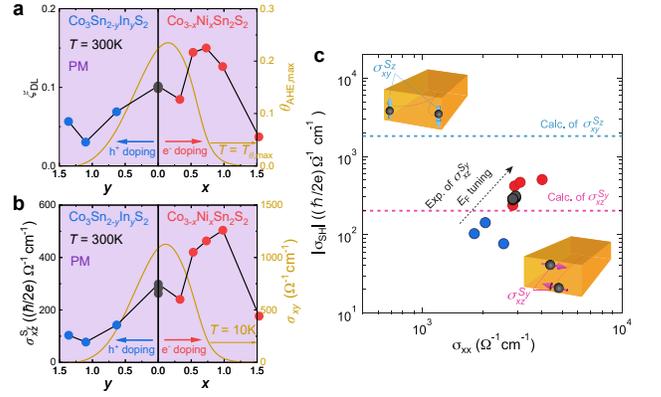}
  \caption{%\textbf{Intercorrelated anomalous Hall and spin Hall conductivity in Co$_{3}$Sn$_2$S$_2$}
  (\textbf{a,b}) Nickel (Ni) composition $x$ and Indium(In) composition $y$ dependence of the damping-like spin Hall efficiency $\xi_\textrm{DL}$ (\textbf{a}) and spin Hall conductivity $\sigma_{xz}^{S_y}$ (\textbf{b}), for paramagnetic shandites based on the harmonic Hall technique, measured at \SI{300}{\kelvin}. Right axes and dark yellow lines show the maximum anomalous Hall angle $\theta_\textrm{AHE,max}$ and anomalous Hall conductivity $\sigma_{xy}$ measured at low temperature. (\textbf{c}) $\sigma_{xz}^{S_y}$ against the longitudinal conductivity $\sigma_{xx}$ for all the samples. The calculated spin Hall conductivity maxima for $\sigma_{xz}^{S_y}$ and $\sigma_{xy}^{S_z}$ are indicated by the dashed lines.}
  \label{fig:Doping-SOT}
\end{center}
\end{figure}

We now extend the harmonic Hall SOT quantification at $T=\SI{300}{\kelvin}$ to Ni-substituted and In-substituted CSS/Cu/CoFeB trilayers. Contrary to $\sigma_{xy}$ that is maximized for undoped FM-CSS, both $\xi_\textrm{DL}$ and $\sigma_{xz}^{S_y}$ [Figs.~\ref{fig:Doping-SOT}(a) and \ref{fig:Doping-SOT}(b)] exhibit a pronounce peak for electron-doped PM shandites. The highest $\xi_\textrm{DL} = +0.15$ is achieved for PM-Co$_{2.25}$Ni$_{0.75}$Sn$_2$S$_2$ ($x \sim 0.75$), showing a large controllability of one order of magnitude by the composition variation. Taking into account the conductivity enhancement on increasing $x$ [Fig.~\ref{fig:singlelayer}(f)], the peak of $\sigma_{xz}^{S_y} \approx 500 (\hbar/2e)\Omega^{-1}\text{cm}^{-1}$ is shifted to Co$_{2.02}$Ni$_{0.98}$Sn$_2$S$_2$ ($x \sim 1.0$). Both $\xi_\textrm{DL}$ and $\sigma_{xz}^{S_y}$ decrease dramatically upon increasing Ni composition to $x \sim 1.5$ or introducing In for hole doping. A direct comparison of the harmonic Hall results for CNSS($x \sim 1.0$)/Cu/CoFeB trilayer and CSIS($y~0.6$)/Cu/CoFeB trilayer is shown in Fig.~S3 in the Supplemental Material \cite{Supplementary}. Compared to the $\theta_\textrm{AHE,max}$ and AHC ($\sigma_{xy}$) peaks near $x\sim0$ for FM-CSS [Right axes and dark yellow lines in Figs.~\ref{fig:Doping-SOT}(a) and \ref{fig:Doping-SOT}(b); reproduced from Figs.~\ref{fig:singlelayer}(g) and \ref{fig:singlelayer}(h)], the observed peak shifts in the composition dependence of $\xi_\textrm{DL}$ and SHC ($\sigma_{xz}^{S_y}$) for PM-CSS agree with our calculations in Fig.~\ref{fig:schematic}.
Since the conductivity of CNSS is in the "moderately dirty" regime, one may consider the intrinsic contribution dominates the SHC. The observed evolution of SHC against $x$ and $y$ thus mainly reflects the relative position between $E_\textrm{F}$ and a source of spin Berry curvature in momentum space, e.g., a gapped Dirac point. The intercorrelation between the AHE and SHE shown here may constitute a transport signature of Weyl-Dirac topological transition in mWSM and the associated redistribution of the electron's filling \cite{BelopolskiPRL2021,LiuDFPRB2021}.

The experimental $\sigma_{xz}^{S_y}$, spanning over a factor of $\sim5$, is plotted against the longitudinal conductivity $\sigma_{xx}$ and compared with the calculated $\sigma_{xz}^{S_y}$ and $\sigma_{xy}^{S_z}$ maxima [horizontal dashed lines] in Fig.~\ref{fig:Doping-SOT}(c). We consider misoriented CSS/Cu interface and other extrinsic mechanisms \cite{ShenPRL2020} may result in experimental $\sigma_{xz}^{S_y}$ that exceeds the calculated maximum. More interestingly, the highly anisotropic SHC tensor inherit from the CSS kagome lattice may allow further SHC enhancement, provided high-quality shandite films with $\textbf{c}$-axis lying in the film plane can be stabilized.

Compared to AHE which can be easily measured for a slab-shape ferromagnetic sample of any size, %with a finite $\lambda$ (often of the order of a few nanometers),
observation and quantification of SHE is far more challenging, rendering its vast screening laborious and unrealistic. Here, we have demonstrated a strategy where the large intrinsic AHE of a ferromagnetic material may serve as a facile indicator for predicting new paramagnetic compounds with potentially enhanced intrinsic SHE. This rule of thumb is best applied to ferromagnets where contribution from one spin channel dominates its intrinsic AHE and Berry curvature, as exemplified by the half-metallic mWSM Co$_{3}$Sn$_2$S$_2$. Another prototypal mWSM Co$_2$MnGa \cite{Belopolski1278} belonging to the highly tunable full Heusler family \cite{MannaPRX2018} may also work, provided a proper dopant that simultaneously introduces additional electrons and reduces $T_\textrm{C}$ can be identified.
Reciprocally, introducing a ferromagnetic hole dopant into well-established paramagnetic spin Hall materials may lead to the discovery of new ferromagnets with large AHC. This may explain the recent demonstration of large AHC in L1$_2$-ordered CrPt$_3$ compound \cite{Markou2021}. One should however be cautious in view of the metallic nature of these materials and the strong spin-orbit coupling of Pt. As a final remark, this strategy is readily extendable to the material screening for thermoelectric generation via the anomalous Nernst effect \cite{Sakai2020,Asaba_SA2021,Pan2022} and spin Nernst effect \cite{Meyer2017,SNE_Sheng}, provided in addition the Mott relation \cite{PuPRL2008} is satisfied.

%Acknowledgements
\section*{Acknowledgements}
The authors are grateful to T. Sasaki and T. Kubota for their help in the preparation of trilayer heterostructures. This work was supported by JSPS KAKENHI Grant-in-Aid for Scientific Research (S) (JP18H05246), Grant-in-Aid for Early-Career Scientists (Grant No. JP20K15156), Grant-in-Aid for Scientific Research (A) (JP20H00299), Grant-in-Aid for Scientific Research (B) (JP20H01830), and JST CREST (JPMJCR18T2). A.O was supported by GP-Spin at Tohoku University and by JST SPRING (Grant No. JPMJSP2114).

%\section*{Author contributions}
%Y.-C.L., K.F., T.S., and A.T. conceived the idea and planned the study. J.I. and K.F. grew and optimized the shandite films, measured the XRD spectra and the basic electrical transport properties. Y.-C.L. grew the trilayer heterostructures, fabricated the devices, measured the AFM, performed the harmonic and FMR measurements, analyzed the measured data, and postulated the enhancement of SHC in electron-doped CSS. A.O. and K.N. performed effective model calculations and contributed to the theoretical interpretations of the experimental results. Y.-C.L. wrote the manuscript with input from all other authors. All authors contributed to the discussion of results. T.S. and A.T. supervised the study.

%\section*{Competing interests}
%The authors declare no competing interests.

%\section*{Data Availability}
%The data that support the plots within this paper and other findings of this study are available from the corresponding author upon reasonable request.

%\section*{References}
% reference
%\clearpage
%\bibliography{CSS_ref}

\begin{thebibliography}{45}%
\makeatletter
\providecommand \@ifxundefined [1]{%
 \@ifx{#1\undefined}
}%
\providecommand \@ifnum [1]{%
 \ifnum #1\expandafter \@firstoftwo
 \else \expandafter \@secondoftwo
 \fi
}%
\providecommand \@ifx [1]{%
 \ifx #1\expandafter \@firstoftwo
 \else \expandafter \@secondoftwo
 \fi
}%
\providecommand \natexlab [1]{#1}%
\providecommand \enquote  [1]{``#1''}%
\providecommand \bibnamefont  [1]{#1}%
\providecommand \bibfnamefont [1]{#1}%
\providecommand \citenamefont [1]{#1}%
\providecommand \href@noop [0]{\@secondoftwo}%
\providecommand \href [0]{\begingroup \@sanitize@url \@href}%
\providecommand \@href[1]{\@@startlink{#1}\@@href}%
\providecommand \@@href[1]{\endgroup#1\@@endlink}%
\providecommand \@sanitize@url [0]{\catcode `\\12\catcode `\$12\catcode
  `\&12\catcode `\#12\catcode `\^12\catcode `\_12\catcode `\%12\relax}%
\providecommand \@@startlink[1]{}%
\providecommand \@@endlink[0]{}%
\providecommand \url  [0]{\begingroup\@sanitize@url \@url }%
\providecommand \@url [1]{\endgroup\@href {#1}{\urlprefix }}%
\providecommand \urlprefix  [0]{URL }%
\providecommand \Eprint [0]{\href }%
\providecommand \doibase [0]{https://doi.org/}%
\providecommand \selectlanguage [0]{\@gobble}%
\providecommand \bibinfo  [0]{\@secondoftwo}%
\providecommand \bibfield  [0]{\@secondoftwo}%
\providecommand \translation [1]{[#1]}%
\providecommand \BibitemOpen [0]{}%
\providecommand \bibitemStop [0]{}%
\providecommand \bibitemNoStop [0]{.\EOS\space}%
\providecommand \EOS [0]{\spacefactor3000\relax}%
\providecommand \BibitemShut  [1]{\csname bibitem#1\endcsname}%
\let\auto@bib@innerbib\@empty
%</preamble>
\bibitem [{\citenamefont {Berry}(1984)}]{Berry1984}%
  \BibitemOpen
  \bibfield  {author} {\bibinfo {author} {\bibfnamefont {M.~V.}\ \bibnamefont
  {Berry}},\ }\bibfield  {title} {\bibinfo {title} {Quantal phase factors
  accompanying adiabatic changes},\ }\href
  {https://doi.org/10.1098/rspa.1984.0023} {\bibfield  {journal} {\bibinfo
  {journal} {Proceedings of the Royal Society of London. A. Mathematical and
  Physical Sciences}\ }\textbf {\bibinfo {volume} {392}},\ \bibinfo {pages}
  {45} (\bibinfo {year} {1984})}\BibitemShut {NoStop}%
\bibitem [{\citenamefont {Xiao}\ \emph {et~al.}(2010)\citenamefont {Xiao},
  \citenamefont {Chang},\ and\ \citenamefont {Niu}}]{RevModPhys_Berry}%
  \BibitemOpen
  \bibfield  {author} {\bibinfo {author} {\bibfnamefont {D.}~\bibnamefont
  {Xiao}}, \bibinfo {author} {\bibfnamefont {M.-C.}\ \bibnamefont {Chang}},\
  and\ \bibinfo {author} {\bibfnamefont {Q.}~\bibnamefont {Niu}},\ }\bibfield
  {title} {\bibinfo {title} {Berry phase effects on electronic properties},\
  }\href {https://doi.org/10.1103/RevModPhys.82.1959} {\bibfield  {journal}
  {\bibinfo  {journal} {Rev. Mod. Phys.}\ }\textbf {\bibinfo {volume} {82}},\
  \bibinfo {pages} {1959} (\bibinfo {year} {2010})}\BibitemShut {NoStop}%
\bibitem [{\citenamefont {Nagaosa}\ \emph {et~al.}(2010)\citenamefont
  {Nagaosa}, \citenamefont {Sinova}, \citenamefont {Onoda}, \citenamefont
  {MacDonald},\ and\ \citenamefont {Ong}}]{RevModPhys_AHE}%
  \BibitemOpen
  \bibfield  {author} {\bibinfo {author} {\bibfnamefont {N.}~\bibnamefont
  {Nagaosa}}, \bibinfo {author} {\bibfnamefont {J.}~\bibnamefont {Sinova}},
  \bibinfo {author} {\bibfnamefont {S.}~\bibnamefont {Onoda}}, \bibinfo
  {author} {\bibfnamefont {A.~H.}\ \bibnamefont {MacDonald}},\ and\ \bibinfo
  {author} {\bibfnamefont {N.~P.}\ \bibnamefont {Ong}},\ }\bibfield  {title}
  {\bibinfo {title} {Anomalous hall effect},\ }\href
  {https://doi.org/10.1103/RevModPhys.82.1539} {\bibfield  {journal} {\bibinfo
  {journal} {Rev. Mod. Phys.}\ }\textbf {\bibinfo {volume} {82}},\ \bibinfo
  {pages} {1539} (\bibinfo {year} {2010})}\BibitemShut {NoStop}%
\bibitem [{\citenamefont {Sinova}\ \emph {et~al.}(2015)\citenamefont {Sinova},
  \citenamefont {Valenzuela}, \citenamefont {Wunderlich}, \citenamefont
  {Back},\ and\ \citenamefont {Jungwirth}}]{RevModPhys_SHE}%
  \BibitemOpen
  \bibfield  {author} {\bibinfo {author} {\bibfnamefont {J.}~\bibnamefont
  {Sinova}}, \bibinfo {author} {\bibfnamefont {S.~O.}\ \bibnamefont
  {Valenzuela}}, \bibinfo {author} {\bibfnamefont {J.}~\bibnamefont
  {Wunderlich}}, \bibinfo {author} {\bibfnamefont {C.~H.}\ \bibnamefont
  {Back}},\ and\ \bibinfo {author} {\bibfnamefont {T.}~\bibnamefont
  {Jungwirth}},\ }\bibfield  {title} {\bibinfo {title} {Spin hall effects},\
  }\href {https://doi.org/10.1103/RevModPhys.87.1213} {\bibfield  {journal}
  {\bibinfo  {journal} {Rev. Mod. Phys.}\ }\textbf {\bibinfo {volume} {87}},\
  \bibinfo {pages} {1213} (\bibinfo {year} {2015})}\BibitemShut {NoStop}%
\bibitem [{\citenamefont {Manchon}\ \emph {et~al.}(2019)\citenamefont
  {Manchon}, \citenamefont {\ifmmode~\check{Z}\else \v{Z}\fi{}elezn\'y},
  \citenamefont {Miron}, \citenamefont {Jungwirth}, \citenamefont {Sinova},
  \citenamefont {Thiaville}, \citenamefont {Garello},\ and\ \citenamefont
  {Gambardella}}]{Manchon_SOTreview}%
  \BibitemOpen
  \bibfield  {author} {\bibinfo {author} {\bibfnamefont {A.}~\bibnamefont
  {Manchon}}, \bibinfo {author} {\bibfnamefont {J.}~\bibnamefont
  {\ifmmode~\check{Z}\else \v{Z}\fi{}elezn\'y}}, \bibinfo {author}
  {\bibfnamefont {I.~M.}\ \bibnamefont {Miron}}, \bibinfo {author}
  {\bibfnamefont {T.}~\bibnamefont {Jungwirth}}, \bibinfo {author}
  {\bibfnamefont {J.}~\bibnamefont {Sinova}}, \bibinfo {author} {\bibfnamefont
  {A.}~\bibnamefont {Thiaville}}, \bibinfo {author} {\bibfnamefont
  {K.}~\bibnamefont {Garello}},\ and\ \bibinfo {author} {\bibfnamefont
  {P.}~\bibnamefont {Gambardella}},\ }\bibfield  {title} {\bibinfo {title}
  {Current-induced spin-orbit torques in ferromagnetic and antiferromagnetic
  systems},\ }\href {https://doi.org/10.1103/RevModPhys.91.035004} {\bibfield
  {journal} {\bibinfo  {journal} {Rev. Mod. Phys.}\ }\textbf {\bibinfo {volume}
  {91}},\ \bibinfo {pages} {035004} (\bibinfo {year} {2019})}\BibitemShut
  {NoStop}%
\bibitem [{\citenamefont {Bass}\ and\ \citenamefont {Pratt}(2007)}]{Bass_2007}%
  \BibitemOpen
  \bibfield  {author} {\bibinfo {author} {\bibfnamefont {J.}~\bibnamefont
  {Bass}}\ and\ \bibinfo {author} {\bibfnamefont {W.~P.}\ \bibnamefont
  {Pratt}},\ }\bibfield  {title} {\bibinfo {title} {Spin-diffusion lengths in
  metals and alloys, and spin-flipping at metal/metal interfaces: an
  experimentalist's critical review},\ }\href
  {https://doi.org/10.1088/0953-8984/19/18/183201} {\bibfield  {journal}
  {\bibinfo  {journal} {Journal of Physics: Condensed Matter}\ }\textbf
  {\bibinfo {volume} {19}},\ \bibinfo {pages} {183201} (\bibinfo {year}
  {2007})}\BibitemShut {NoStop}%
\bibitem [{\citenamefont {Onoda}\ \emph {et~al.}(2008)\citenamefont {Onoda},
  \citenamefont {Sugimoto},\ and\ \citenamefont {Nagaosa}}]{OnodaPRB2008}%
  \BibitemOpen
  \bibfield  {author} {\bibinfo {author} {\bibfnamefont {S.}~\bibnamefont
  {Onoda}}, \bibinfo {author} {\bibfnamefont {N.}~\bibnamefont {Sugimoto}},\
  and\ \bibinfo {author} {\bibfnamefont {N.}~\bibnamefont {Nagaosa}},\
  }\bibfield  {title} {\bibinfo {title} {Quantum transport theory of anomalous
  electric, thermoelectric, and thermal hall effects in ferromagnets},\ }\href
  {https://doi.org/10.1103/PhysRevB.77.165103} {\bibfield  {journal} {\bibinfo
  {journal} {Phys. Rev. B}\ }\textbf {\bibinfo {volume} {77}},\ \bibinfo
  {pages} {165103} (\bibinfo {year} {2008})}\BibitemShut {NoStop}%
\bibitem [{\citenamefont {Moriya}\ \emph {et~al.}(2022)\citenamefont {Moriya},
  \citenamefont {Musha}, \citenamefont {Haku},\ and\ \citenamefont
  {Ando}}]{Moriya2022}%
  \BibitemOpen
  \bibfield  {author} {\bibinfo {author} {\bibfnamefont {H.}~\bibnamefont
  {Moriya}}, \bibinfo {author} {\bibfnamefont {A.}~\bibnamefont {Musha}},
  \bibinfo {author} {\bibfnamefont {S.}~\bibnamefont {Haku}},\ and\ \bibinfo
  {author} {\bibfnamefont {K.}~\bibnamefont {Ando}},\ }\bibfield  {title}
  {\bibinfo {title} {Observation of the crossover between metallic and
  insulating regimes of the spin hall effect},\ }\href
  {https://doi.org/10.1038/s42005-021-00791-1} {\bibfield  {journal} {\bibinfo
  {journal} {Communications Physics}\ }\textbf {\bibinfo {volume} {5}},\
  \bibinfo {pages} {12} (\bibinfo {year} {2022})}\BibitemShut {NoStop}%
\bibitem [{\citenamefont {Omori}\ \emph {et~al.}(2019)\citenamefont {Omori},
  \citenamefont {Sagasta}, \citenamefont {Niimi}, \citenamefont {Gradhand},
  \citenamefont {Hueso}, \citenamefont {Casanova},\ and\ \citenamefont
  {Otani}}]{Omori_PRB_2019}%
  \BibitemOpen
  \bibfield  {author} {\bibinfo {author} {\bibfnamefont {Y.}~\bibnamefont
  {Omori}}, \bibinfo {author} {\bibfnamefont {E.}~\bibnamefont {Sagasta}},
  \bibinfo {author} {\bibfnamefont {Y.}~\bibnamefont {Niimi}}, \bibinfo
  {author} {\bibfnamefont {M.}~\bibnamefont {Gradhand}}, \bibinfo {author}
  {\bibfnamefont {L.~E.}\ \bibnamefont {Hueso}}, \bibinfo {author}
  {\bibfnamefont {F.}~\bibnamefont {Casanova}},\ and\ \bibinfo {author}
  {\bibfnamefont {Y.}~\bibnamefont {Otani}},\ }\bibfield  {title} {\bibinfo
  {title} {Relation between spin hall effect and anomalous hall effect in $3d$
  ferromagnetic metals},\ }\href {https://doi.org/10.1103/PhysRevB.99.014403}
  {\bibfield  {journal} {\bibinfo  {journal} {Phys. Rev. B}\ }\textbf {\bibinfo
  {volume} {99}},\ \bibinfo {pages} {014403} (\bibinfo {year}
  {2019})}\BibitemShut {NoStop}%
\bibitem [{\citenamefont {Vafek}\ and\ \citenamefont
  {Vishwanath}(2014)}]{VafekRev2014}%
  \BibitemOpen
  \bibfield  {author} {\bibinfo {author} {\bibfnamefont {O.}~\bibnamefont
  {Vafek}}\ and\ \bibinfo {author} {\bibfnamefont {A.}~\bibnamefont
  {Vishwanath}},\ }\bibfield  {title} {\bibinfo {title} {Dirac fermions in
  solids: From high-tc cuprates and graphene to topological insulators and weyl
  semimetals},\ }\href
  {https://doi.org/10.1146/annurev-conmatphys-031113-133841} {\bibfield
  {journal} {\bibinfo  {journal} {Annual Review of Condensed Matter Physics}\
  }\textbf {\bibinfo {volume} {5}},\ \bibinfo {pages} {83} (\bibinfo {year}
  {2014})}\BibitemShut {NoStop}%
\bibitem [{\citenamefont {Yan}\ and\ \citenamefont
  {Felser}(2017)}]{YanRev2017}%
  \BibitemOpen
  \bibfield  {author} {\bibinfo {author} {\bibfnamefont {B.}~\bibnamefont
  {Yan}}\ and\ \bibinfo {author} {\bibfnamefont {C.}~\bibnamefont {Felser}},\
  }\bibfield  {title} {\bibinfo {title} {Topological materials: Weyl
  semimetals},\ }\href
  {https://doi.org/10.1146/annurev-conmatphys-031016-025458} {\bibfield
  {journal} {\bibinfo  {journal} {Annual Review of Condensed Matter Physics}\
  }\textbf {\bibinfo {volume} {8}},\ \bibinfo {pages} {337} (\bibinfo {year}
  {2017})}\BibitemShut {NoStop}%
\bibitem [{\citenamefont {Armitage}\ \emph {et~al.}(2018)\citenamefont
  {Armitage}, \citenamefont {Mele},\ and\ \citenamefont
  {Vishwanath}}]{RevModPhys_WSM}%
  \BibitemOpen
  \bibfield  {author} {\bibinfo {author} {\bibfnamefont {N.~P.}\ \bibnamefont
  {Armitage}}, \bibinfo {author} {\bibfnamefont {E.~J.}\ \bibnamefont {Mele}},\
  and\ \bibinfo {author} {\bibfnamefont {A.}~\bibnamefont {Vishwanath}},\
  }\bibfield  {title} {\bibinfo {title} {Weyl and dirac semimetals in
  three-dimensional solids},\ }\href
  {https://doi.org/10.1103/RevModPhys.90.015001} {\bibfield  {journal}
  {\bibinfo  {journal} {Rev. Mod. Phys.}\ }\textbf {\bibinfo {volume} {90}},\
  \bibinfo {pages} {015001} (\bibinfo {year} {2018})}\BibitemShut {NoStop}%
\bibitem [{\citenamefont {Nagaosa}\ \emph {et~al.}(2020)\citenamefont
  {Nagaosa}, \citenamefont {Morimoto},\ and\ \citenamefont
  {Tokura}}]{Nagaosa_NRM2020}%
  \BibitemOpen
  \bibfield  {author} {\bibinfo {author} {\bibfnamefont {N.}~\bibnamefont
  {Nagaosa}}, \bibinfo {author} {\bibfnamefont {T.}~\bibnamefont {Morimoto}},\
  and\ \bibinfo {author} {\bibfnamefont {Y.}~\bibnamefont {Tokura}},\
  }\bibfield  {title} {\bibinfo {title} {Transport, magnetic and optical
  properties of weyl materials},\ }\href
  {https://doi.org/10.1038/s41578-020-0208-y} {\bibfield  {journal} {\bibinfo
  {journal} {Nature Reviews Materials}\ }\textbf {\bibinfo {volume} {5}},\
  \bibinfo {pages} {621} (\bibinfo {year} {2020})}\BibitemShut {NoStop}%
\bibitem [{\citenamefont {Liu}\ \emph {et~al.}(2019)\citenamefont {Liu},
  \citenamefont {Liang}, \citenamefont {Liu}, \citenamefont {Xu}, \citenamefont
  {Li}, \citenamefont {Chen}, \citenamefont {Pei}, \citenamefont {Shi},
  \citenamefont {Mo}, \citenamefont {Dudin}, \citenamefont {Kim}, \citenamefont
  {Cacho}, \citenamefont {Li}, \citenamefont {Sun}, \citenamefont {Yang},
  \citenamefont {Liu}, \citenamefont {Parkin}, \citenamefont {Felser},\ and\
  \citenamefont {Chen}}]{Liu1282}%
  \BibitemOpen
  \bibfield  {author} {\bibinfo {author} {\bibfnamefont {D.~F.}\ \bibnamefont
  {Liu}}, \bibinfo {author} {\bibfnamefont {A.~J.}\ \bibnamefont {Liang}},
  \bibinfo {author} {\bibfnamefont {E.~K.}\ \bibnamefont {Liu}}, \bibinfo
  {author} {\bibfnamefont {Q.~N.}\ \bibnamefont {Xu}}, \bibinfo {author}
  {\bibfnamefont {Y.~W.}\ \bibnamefont {Li}}, \bibinfo {author} {\bibfnamefont
  {C.}~\bibnamefont {Chen}}, \bibinfo {author} {\bibfnamefont {D.}~\bibnamefont
  {Pei}}, \bibinfo {author} {\bibfnamefont {W.~J.}\ \bibnamefont {Shi}},
  \bibinfo {author} {\bibfnamefont {S.~K.}\ \bibnamefont {Mo}}, \bibinfo
  {author} {\bibfnamefont {P.}~\bibnamefont {Dudin}}, \bibinfo {author}
  {\bibfnamefont {T.}~\bibnamefont {Kim}}, \bibinfo {author} {\bibfnamefont
  {C.}~\bibnamefont {Cacho}}, \bibinfo {author} {\bibfnamefont
  {G.}~\bibnamefont {Li}}, \bibinfo {author} {\bibfnamefont {Y.}~\bibnamefont
  {Sun}}, \bibinfo {author} {\bibfnamefont {L.~X.}\ \bibnamefont {Yang}},
  \bibinfo {author} {\bibfnamefont {Z.~K.}\ \bibnamefont {Liu}}, \bibinfo
  {author} {\bibfnamefont {S.~S.~P.}\ \bibnamefont {Parkin}}, \bibinfo {author}
  {\bibfnamefont {C.}~\bibnamefont {Felser}},\ and\ \bibinfo {author}
  {\bibfnamefont {Y.~L.}\ \bibnamefont {Chen}},\ }\bibfield  {title} {\bibinfo
  {title} {Magnetic weyl semimetal phase in a kagom{\'e} crystal},\ }\href
  {https://doi.org/10.1126/science.aav2873} {\bibfield  {journal} {\bibinfo
  {journal} {Science}\ }\textbf {\bibinfo {volume} {365}},\ \bibinfo {pages}
  {1282} (\bibinfo {year} {2019})}\BibitemShut {NoStop}%
\bibitem [{\citenamefont {Morali}\ \emph {et~al.}(2019)\citenamefont {Morali},
  \citenamefont {Batabyal}, \citenamefont {Nag}, \citenamefont {Liu},
  \citenamefont {Xu}, \citenamefont {Sun}, \citenamefont {Yan}, \citenamefont
  {Felser}, \citenamefont {Avraham},\ and\ \citenamefont
  {Beidenkopf}}]{Morali1286}%
  \BibitemOpen
  \bibfield  {author} {\bibinfo {author} {\bibfnamefont {N.}~\bibnamefont
  {Morali}}, \bibinfo {author} {\bibfnamefont {R.}~\bibnamefont {Batabyal}},
  \bibinfo {author} {\bibfnamefont {P.~K.}\ \bibnamefont {Nag}}, \bibinfo
  {author} {\bibfnamefont {E.}~\bibnamefont {Liu}}, \bibinfo {author}
  {\bibfnamefont {Q.}~\bibnamefont {Xu}}, \bibinfo {author} {\bibfnamefont
  {Y.}~\bibnamefont {Sun}}, \bibinfo {author} {\bibfnamefont {B.}~\bibnamefont
  {Yan}}, \bibinfo {author} {\bibfnamefont {C.}~\bibnamefont {Felser}},
  \bibinfo {author} {\bibfnamefont {N.}~\bibnamefont {Avraham}},\ and\ \bibinfo
  {author} {\bibfnamefont {H.}~\bibnamefont {Beidenkopf}},\ }\bibfield  {title}
  {\bibinfo {title} {Fermi-arc diversity on surface terminations of the
  magnetic weyl semimetal co3sn2s2},\ }\href
  {https://doi.org/10.1126/science.aav2334} {\bibfield  {journal} {\bibinfo
  {journal} {Science}\ }\textbf {\bibinfo {volume} {365}},\ \bibinfo {pages}
  {1286} (\bibinfo {year} {2019})}\BibitemShut {NoStop}%
\bibitem [{\citenamefont {Liu}\ \emph {et~al.}(2018)\citenamefont {Liu},
  \citenamefont {Sun}, \citenamefont {Kumar}, \citenamefont {Muechler},
  \citenamefont {Sun}, \citenamefont {Jiao}, \citenamefont {Yang},
  \citenamefont {Liu}, \citenamefont {Liang}, \citenamefont {Xu}, \citenamefont
  {Kroder}, \citenamefont {S{\"u}{\ss}}, \citenamefont {Borrmann},
  \citenamefont {Shekhar}, \citenamefont {Wang}, \citenamefont {Xi},
  \citenamefont {Wang}, \citenamefont {Schnelle}, \citenamefont {Wirth},
  \citenamefont {Chen}, \citenamefont {Goennenwein},\ and\ \citenamefont
  {Felser}}]{Liu2018}%
  \BibitemOpen
  \bibfield  {author} {\bibinfo {author} {\bibfnamefont {E.}~\bibnamefont
  {Liu}}, \bibinfo {author} {\bibfnamefont {Y.}~\bibnamefont {Sun}}, \bibinfo
  {author} {\bibfnamefont {N.}~\bibnamefont {Kumar}}, \bibinfo {author}
  {\bibfnamefont {L.}~\bibnamefont {Muechler}}, \bibinfo {author}
  {\bibfnamefont {A.}~\bibnamefont {Sun}}, \bibinfo {author} {\bibfnamefont
  {L.}~\bibnamefont {Jiao}}, \bibinfo {author} {\bibfnamefont {S.-Y.}\
  \bibnamefont {Yang}}, \bibinfo {author} {\bibfnamefont {D.}~\bibnamefont
  {Liu}}, \bibinfo {author} {\bibfnamefont {A.}~\bibnamefont {Liang}}, \bibinfo
  {author} {\bibfnamefont {Q.}~\bibnamefont {Xu}}, \bibinfo {author}
  {\bibfnamefont {J.}~\bibnamefont {Kroder}}, \bibinfo {author} {\bibfnamefont
  {V.}~\bibnamefont {S{\"u}{\ss}}}, \bibinfo {author} {\bibfnamefont
  {H.}~\bibnamefont {Borrmann}}, \bibinfo {author} {\bibfnamefont
  {C.}~\bibnamefont {Shekhar}}, \bibinfo {author} {\bibfnamefont
  {Z.}~\bibnamefont {Wang}}, \bibinfo {author} {\bibfnamefont {C.}~\bibnamefont
  {Xi}}, \bibinfo {author} {\bibfnamefont {W.}~\bibnamefont {Wang}}, \bibinfo
  {author} {\bibfnamefont {W.}~\bibnamefont {Schnelle}}, \bibinfo {author}
  {\bibfnamefont {S.}~\bibnamefont {Wirth}}, \bibinfo {author} {\bibfnamefont
  {Y.}~\bibnamefont {Chen}}, \bibinfo {author} {\bibfnamefont {S.~T.~B.}\
  \bibnamefont {Goennenwein}},\ and\ \bibinfo {author} {\bibfnamefont
  {C.}~\bibnamefont {Felser}},\ }\bibfield  {title} {\bibinfo {title} {Giant
  anomalous hall effect in a ferromagnetic kagome-lattice semimetal},\ }\href
  {https://doi.org/10.1038/s41567-018-0234-5} {\bibfield  {journal} {\bibinfo
  {journal} {Nature Physics}\ }\textbf {\bibinfo {volume} {14}},\ \bibinfo
  {pages} {1125} (\bibinfo {year} {2018})}\BibitemShut {NoStop}%
\bibitem [{\citenamefont {Wang}\ \emph {et~al.}(2018)\citenamefont {Wang},
  \citenamefont {Xu}, \citenamefont {Lou}, \citenamefont {Liu}, \citenamefont
  {Li}, \citenamefont {Huang}, \citenamefont {Shen}, \citenamefont {Weng},
  \citenamefont {Wang},\ and\ \citenamefont {Lei}}]{Wang2018}%
  \BibitemOpen
  \bibfield  {author} {\bibinfo {author} {\bibfnamefont {Q.}~\bibnamefont
  {Wang}}, \bibinfo {author} {\bibfnamefont {Y.}~\bibnamefont {Xu}}, \bibinfo
  {author} {\bibfnamefont {R.}~\bibnamefont {Lou}}, \bibinfo {author}
  {\bibfnamefont {Z.}~\bibnamefont {Liu}}, \bibinfo {author} {\bibfnamefont
  {M.}~\bibnamefont {Li}}, \bibinfo {author} {\bibfnamefont {Y.}~\bibnamefont
  {Huang}}, \bibinfo {author} {\bibfnamefont {D.}~\bibnamefont {Shen}},
  \bibinfo {author} {\bibfnamefont {H.}~\bibnamefont {Weng}}, \bibinfo {author}
  {\bibfnamefont {S.}~\bibnamefont {Wang}},\ and\ \bibinfo {author}
  {\bibfnamefont {H.}~\bibnamefont {Lei}},\ }\bibfield  {title} {\bibinfo
  {title} {Large intrinsic anomalous hall effect in half-metallic ferromagnet
  co3sn2s2 with magnetic weyl fermions},\ }\href
  {https://doi.org/10.1038/s41467-018-06088-2} {\bibfield  {journal} {\bibinfo
  {journal} {Nature Communications}\ }\textbf {\bibinfo {volume} {9}},\
  \bibinfo {pages} {3681} (\bibinfo {year} {2018})}\BibitemShut {NoStop}%
\bibitem [{\citenamefont {Li}\ \emph {et~al.}(2019)\citenamefont {Li},
  \citenamefont {Xu}, \citenamefont {Shi}, \citenamefont {Fu}, \citenamefont
  {Jiao}, \citenamefont {Kamminga}, \citenamefont {Yu}, \citenamefont
  {T{\"u}ys{\"u}z}, \citenamefont {Kumar}, \citenamefont {S{\"u}{\ss}},
  \citenamefont {Saha}, \citenamefont {Srivastava}, \citenamefont {Wirth},
  \citenamefont {Auffermann}, \citenamefont {Gooth}, \citenamefont {Parkin},
  \citenamefont {Sun}, \citenamefont {Liu},\ and\ \citenamefont
  {Felser}}]{Li_CSS_oxydation}%
  \BibitemOpen
  \bibfield  {author} {\bibinfo {author} {\bibfnamefont {G.}~\bibnamefont
  {Li}}, \bibinfo {author} {\bibfnamefont {Q.}~\bibnamefont {Xu}}, \bibinfo
  {author} {\bibfnamefont {W.}~\bibnamefont {Shi}}, \bibinfo {author}
  {\bibfnamefont {C.}~\bibnamefont {Fu}}, \bibinfo {author} {\bibfnamefont
  {L.}~\bibnamefont {Jiao}}, \bibinfo {author} {\bibfnamefont {M.~E.}\
  \bibnamefont {Kamminga}}, \bibinfo {author} {\bibfnamefont {M.}~\bibnamefont
  {Yu}}, \bibinfo {author} {\bibfnamefont {H.}~\bibnamefont {T{\"u}ys{\"u}z}},
  \bibinfo {author} {\bibfnamefont {N.}~\bibnamefont {Kumar}}, \bibinfo
  {author} {\bibfnamefont {V.}~\bibnamefont {S{\"u}{\ss}}}, \bibinfo {author}
  {\bibfnamefont {R.}~\bibnamefont {Saha}}, \bibinfo {author} {\bibfnamefont
  {A.~K.}\ \bibnamefont {Srivastava}}, \bibinfo {author} {\bibfnamefont
  {S.}~\bibnamefont {Wirth}}, \bibinfo {author} {\bibfnamefont
  {G.}~\bibnamefont {Auffermann}}, \bibinfo {author} {\bibfnamefont
  {J.}~\bibnamefont {Gooth}}, \bibinfo {author} {\bibfnamefont
  {S.}~\bibnamefont {Parkin}}, \bibinfo {author} {\bibfnamefont
  {Y.}~\bibnamefont {Sun}}, \bibinfo {author} {\bibfnamefont {E.}~\bibnamefont
  {Liu}},\ and\ \bibinfo {author} {\bibfnamefont {C.}~\bibnamefont {Felser}},\
  }\bibfield  {title} {\bibinfo {title} {Surface states in bulk single crystal
  of topological semimetal co3sn2s2 toward water oxidation},\ }\href
  {https://doi.org/10.1126/sciadv.aaw9867} {\bibfield  {journal} {\bibinfo
  {journal} {Science Advances}\ }\textbf {\bibinfo {volume} {5}},\ \bibinfo
  {pages} {eaaw9867} (\bibinfo {year} {2019})}\BibitemShut {NoStop}%
\bibitem [{\citenamefont {Ozawa}\ and\ \citenamefont
  {Nomura}(2019)}]{OzawaJPSJ2019}%
  \BibitemOpen
  \bibfield  {author} {\bibinfo {author} {\bibfnamefont {A.}~\bibnamefont
  {Ozawa}}\ and\ \bibinfo {author} {\bibfnamefont {K.}~\bibnamefont {Nomura}},\
  }\bibfield  {title} {\bibinfo {title} {Two-orbital effective model for
  magnetic weyl semimetal in kagome-lattice shandite},\ }\href
  {https://doi.org/10.7566/JPSJ.88.123703} {\bibfield  {journal} {\bibinfo
  {journal} {Journal of the Physical Society of Japan}\ }\textbf {\bibinfo
  {volume} {88}},\ \bibinfo {pages} {123703} (\bibinfo {year}
  {2019})}\BibitemShut {NoStop}%
\bibitem [{Sup()}]{Supplementary}%
  \BibitemOpen
  \href@noop {} {}\bibinfo {note} {See Supplemental Material at [URL] for
  detailed sample preparation, magneto-transport, harmonic Hall technique
  analysis, spin-torque ferromagnetic resonance analysis, and tight-binding
  effective-band model calculations.}\BibitemShut {Stop}%
\bibitem [{\citenamefont {Belopolski}\ \emph {et~al.}(2021)\citenamefont
  {Belopolski}, \citenamefont {Cochran}, \citenamefont {Liu}, \citenamefont
  {Cheng}, \citenamefont {Yang}, \citenamefont {Guguchia}, \citenamefont
  {Tsirkin}, \citenamefont {Yin}, \citenamefont {Vir}, \citenamefont {Thakur},
  \citenamefont {Zhang}, \citenamefont {Zhang}, \citenamefont {Kaznatcheev},
  \citenamefont {Cheng}, \citenamefont {Chang}, \citenamefont {Multer},
  \citenamefont {Shumiya}, \citenamefont {Litskevich}, \citenamefont {Vescovo},
  \citenamefont {Kim}, \citenamefont {Cacho}, \citenamefont {Yao},
  \citenamefont {Felser}, \citenamefont {Neupert},\ and\ \citenamefont
  {Hasan}}]{BelopolskiPRL2021}%
  \BibitemOpen
  \bibfield  {author} {\bibinfo {author} {\bibfnamefont {I.}~\bibnamefont
  {Belopolski}}, \bibinfo {author} {\bibfnamefont {T.~A.}\ \bibnamefont
  {Cochran}}, \bibinfo {author} {\bibfnamefont {X.}~\bibnamefont {Liu}},
  \bibinfo {author} {\bibfnamefont {Z.-J.}\ \bibnamefont {Cheng}}, \bibinfo
  {author} {\bibfnamefont {X.~P.}\ \bibnamefont {Yang}}, \bibinfo {author}
  {\bibfnamefont {Z.}~\bibnamefont {Guguchia}}, \bibinfo {author}
  {\bibfnamefont {S.~S.}\ \bibnamefont {Tsirkin}}, \bibinfo {author}
  {\bibfnamefont {J.-X.}\ \bibnamefont {Yin}}, \bibinfo {author} {\bibfnamefont
  {P.}~\bibnamefont {Vir}}, \bibinfo {author} {\bibfnamefont {G.~S.}\
  \bibnamefont {Thakur}}, \bibinfo {author} {\bibfnamefont {S.~S.}\
  \bibnamefont {Zhang}}, \bibinfo {author} {\bibfnamefont {J.}~\bibnamefont
  {Zhang}}, \bibinfo {author} {\bibfnamefont {K.}~\bibnamefont {Kaznatcheev}},
  \bibinfo {author} {\bibfnamefont {G.}~\bibnamefont {Cheng}}, \bibinfo
  {author} {\bibfnamefont {G.}~\bibnamefont {Chang}}, \bibinfo {author}
  {\bibfnamefont {D.}~\bibnamefont {Multer}}, \bibinfo {author} {\bibfnamefont
  {N.}~\bibnamefont {Shumiya}}, \bibinfo {author} {\bibfnamefont
  {M.}~\bibnamefont {Litskevich}}, \bibinfo {author} {\bibfnamefont
  {E.}~\bibnamefont {Vescovo}}, \bibinfo {author} {\bibfnamefont {T.~K.}\
  \bibnamefont {Kim}}, \bibinfo {author} {\bibfnamefont {C.}~\bibnamefont
  {Cacho}}, \bibinfo {author} {\bibfnamefont {N.}~\bibnamefont {Yao}}, \bibinfo
  {author} {\bibfnamefont {C.}~\bibnamefont {Felser}}, \bibinfo {author}
  {\bibfnamefont {T.}~\bibnamefont {Neupert}},\ and\ \bibinfo {author}
  {\bibfnamefont {M.~Z.}\ \bibnamefont {Hasan}},\ }\bibfield  {title} {\bibinfo
  {title} {Signatures of weyl fermion annihilation in a correlated kagome
  magnet},\ }\href {https://doi.org/10.1103/PhysRevLett.127.256403} {\bibfield
  {journal} {\bibinfo  {journal} {Phys. Rev. Lett.}\ }\textbf {\bibinfo
  {volume} {127}},\ \bibinfo {pages} {256403} (\bibinfo {year}
  {2021})}\BibitemShut {NoStop}%
\bibitem [{\citenamefont {Liu}\ \emph {et~al.}(2021)\citenamefont {Liu},
  \citenamefont {Xu}, \citenamefont {Liu}, \citenamefont {Shen}, \citenamefont
  {Le}, \citenamefont {Li}, \citenamefont {Pei}, \citenamefont {Liang},
  \citenamefont {Dudin}, \citenamefont {Kim}, \citenamefont {Cacho},
  \citenamefont {Xu}, \citenamefont {Sun}, \citenamefont {Yang}, \citenamefont
  {Liu}, \citenamefont {Felser}, \citenamefont {Parkin},\ and\ \citenamefont
  {Chen}}]{LiuDFPRB2021}%
  \BibitemOpen
  \bibfield  {author} {\bibinfo {author} {\bibfnamefont {D.~F.}\ \bibnamefont
  {Liu}}, \bibinfo {author} {\bibfnamefont {Q.~N.}\ \bibnamefont {Xu}},
  \bibinfo {author} {\bibfnamefont {E.~K.}\ \bibnamefont {Liu}}, \bibinfo
  {author} {\bibfnamefont {J.~L.}\ \bibnamefont {Shen}}, \bibinfo {author}
  {\bibfnamefont {C.~C.}\ \bibnamefont {Le}}, \bibinfo {author} {\bibfnamefont
  {Y.~W.}\ \bibnamefont {Li}}, \bibinfo {author} {\bibfnamefont
  {D.}~\bibnamefont {Pei}}, \bibinfo {author} {\bibfnamefont {A.~J.}\
  \bibnamefont {Liang}}, \bibinfo {author} {\bibfnamefont {P.}~\bibnamefont
  {Dudin}}, \bibinfo {author} {\bibfnamefont {T.~K.}\ \bibnamefont {Kim}},
  \bibinfo {author} {\bibfnamefont {C.}~\bibnamefont {Cacho}}, \bibinfo
  {author} {\bibfnamefont {Y.~F.}\ \bibnamefont {Xu}}, \bibinfo {author}
  {\bibfnamefont {Y.}~\bibnamefont {Sun}}, \bibinfo {author} {\bibfnamefont
  {L.~X.}\ \bibnamefont {Yang}}, \bibinfo {author} {\bibfnamefont {Z.~K.}\
  \bibnamefont {Liu}}, \bibinfo {author} {\bibfnamefont {C.}~\bibnamefont
  {Felser}}, \bibinfo {author} {\bibfnamefont {S.~S.~P.}\ \bibnamefont
  {Parkin}},\ and\ \bibinfo {author} {\bibfnamefont {Y.~L.}\ \bibnamefont
  {Chen}},\ }\bibfield  {title} {\bibinfo {title} {Topological phase transition
  in a magnetic weyl semimetal},\ }\href
  {https://doi.org/10.1103/PhysRevB.104.205140} {\bibfield  {journal} {\bibinfo
   {journal} {Phys. Rev. B}\ }\textbf {\bibinfo {volume} {104}},\ \bibinfo
  {pages} {205140} (\bibinfo {year} {2021})}\BibitemShut {NoStop}%
\bibitem [{\citenamefont {Guo}\ \emph {et~al.}(2008)\citenamefont {Guo},
  \citenamefont {Murakami}, \citenamefont {Chen},\ and\ \citenamefont
  {Nagaosa}}]{GuoPRL2008}%
  \BibitemOpen
  \bibfield  {author} {\bibinfo {author} {\bibfnamefont {G.~Y.}\ \bibnamefont
  {Guo}}, \bibinfo {author} {\bibfnamefont {S.}~\bibnamefont {Murakami}},
  \bibinfo {author} {\bibfnamefont {T.-W.}\ \bibnamefont {Chen}},\ and\
  \bibinfo {author} {\bibfnamefont {N.}~\bibnamefont {Nagaosa}},\ }\bibfield
  {title} {\bibinfo {title} {Intrinsic spin hall effect in platinum:
  First-principles calculations},\ }\href
  {https://doi.org/10.1103/PhysRevLett.100.096401} {\bibfield  {journal}
  {\bibinfo  {journal} {Phys. Rev. Lett.}\ }\textbf {\bibinfo {volume} {100}},\
  \bibinfo {pages} {096401} (\bibinfo {year} {2008})}\BibitemShut {NoStop}%
\bibitem [{\citenamefont {Hoffmann}(2013)}]{Hoffmann}%
  \BibitemOpen
  \bibfield  {author} {\bibinfo {author} {\bibfnamefont {A.}~\bibnamefont
  {Hoffmann}},\ }\bibfield  {title} {\bibinfo {title} {Spin hall effects in
  metals},\ }\href {https://doi.org/10.1109/TMAG.2013.2262947} {\bibfield
  {journal} {\bibinfo  {journal} {IEEE Transactions on Magnetics}\ }\textbf
  {\bibinfo {volume} {49}},\ \bibinfo {pages} {5172} (\bibinfo {year}
  {2013})}\BibitemShut {NoStop}%
\bibitem [{\citenamefont {Ye}\ \emph {et~al.}(2018)\citenamefont {Ye},
  \citenamefont {Kang}, \citenamefont {Liu}, \citenamefont {von Cube},
  \citenamefont {Wicker}, \citenamefont {Suzuki}, \citenamefont {Jozwiak},
  \citenamefont {Bostwick}, \citenamefont {Rotenberg}, \citenamefont {Bell},
  \citenamefont {Fu}, \citenamefont {Comin},\ and\ \citenamefont
  {Checkelsky}}]{YeNature2018}%
  \BibitemOpen
  \bibfield  {author} {\bibinfo {author} {\bibfnamefont {L.}~\bibnamefont
  {Ye}}, \bibinfo {author} {\bibfnamefont {M.}~\bibnamefont {Kang}}, \bibinfo
  {author} {\bibfnamefont {J.}~\bibnamefont {Liu}}, \bibinfo {author}
  {\bibfnamefont {F.}~\bibnamefont {von Cube}}, \bibinfo {author}
  {\bibfnamefont {C.~R.}\ \bibnamefont {Wicker}}, \bibinfo {author}
  {\bibfnamefont {T.}~\bibnamefont {Suzuki}}, \bibinfo {author} {\bibfnamefont
  {C.}~\bibnamefont {Jozwiak}}, \bibinfo {author} {\bibfnamefont
  {A.}~\bibnamefont {Bostwick}}, \bibinfo {author} {\bibfnamefont
  {E.}~\bibnamefont {Rotenberg}}, \bibinfo {author} {\bibfnamefont {D.~C.}\
  \bibnamefont {Bell}}, \bibinfo {author} {\bibfnamefont {L.}~\bibnamefont
  {Fu}}, \bibinfo {author} {\bibfnamefont {R.}~\bibnamefont {Comin}},\ and\
  \bibinfo {author} {\bibfnamefont {J.~G.}\ \bibnamefont {Checkelsky}},\
  }\bibfield  {title} {\bibinfo {title} {Massive dirac fermions in a
  ferromagnetic kagome metal},\ }\href {https://doi.org/10.1038/nature25987}
  {\bibfield  {journal} {\bibinfo  {journal} {Nature}\ }\textbf {\bibinfo
  {volume} {555}},\ \bibinfo {pages} {638} (\bibinfo {year}
  {2018})}\BibitemShut {NoStop}%
\bibitem [{\citenamefont {Fujiwara}\ \emph {et~al.}(2019)\citenamefont
  {Fujiwara}, \citenamefont {Ikeda}, \citenamefont {Shiogai}, \citenamefont
  {Seki}, \citenamefont {Takanashi},\ and\ \citenamefont
  {Tsukazaki}}]{Fujiwara_2019}%
  \BibitemOpen
  \bibfield  {author} {\bibinfo {author} {\bibfnamefont {K.}~\bibnamefont
  {Fujiwara}}, \bibinfo {author} {\bibfnamefont {J.}~\bibnamefont {Ikeda}},
  \bibinfo {author} {\bibfnamefont {J.}~\bibnamefont {Shiogai}}, \bibinfo
  {author} {\bibfnamefont {T.}~\bibnamefont {Seki}}, \bibinfo {author}
  {\bibfnamefont {K.}~\bibnamefont {Takanashi}},\ and\ \bibinfo {author}
  {\bibfnamefont {A.}~\bibnamefont {Tsukazaki}},\ }\bibfield  {title} {\bibinfo
  {title} {Ferromagnetic co3sn2s2 thin films fabricated by co-sputtering},\
  }\href {https://doi.org/10.7567/1347-4065/ab12ff} {\bibfield  {journal}
  {\bibinfo  {journal} {Japanese Journal of Applied Physics}\ }\textbf
  {\bibinfo {volume} {58}},\ \bibinfo {pages} {050912} (\bibinfo {year}
  {2019})}\BibitemShut {NoStop}%
\bibitem [{\citenamefont {Ikeda}\ \emph
  {et~al.}(2021{\natexlab{a}})\citenamefont {Ikeda}, \citenamefont {Fujiwara},
  \citenamefont {Shiogai}, \citenamefont {Seki}, \citenamefont {Nomura},
  \citenamefont {Takanashi},\ and\ \citenamefont
  {Tsukazaki}}]{Ikeda2021CommMater}%
  \BibitemOpen
  \bibfield  {author} {\bibinfo {author} {\bibfnamefont {J.}~\bibnamefont
  {Ikeda}}, \bibinfo {author} {\bibfnamefont {K.}~\bibnamefont {Fujiwara}},
  \bibinfo {author} {\bibfnamefont {J.}~\bibnamefont {Shiogai}}, \bibinfo
  {author} {\bibfnamefont {T.}~\bibnamefont {Seki}}, \bibinfo {author}
  {\bibfnamefont {K.}~\bibnamefont {Nomura}}, \bibinfo {author} {\bibfnamefont
  {K.}~\bibnamefont {Takanashi}},\ and\ \bibinfo {author} {\bibfnamefont
  {A.}~\bibnamefont {Tsukazaki}},\ }\bibfield  {title} {\bibinfo {title}
  {Critical thickness for the emergence of weyl features in co3sn2s2 thin
  films},\ }\href {https://doi.org/10.1038/s43246-021-00122-5} {\bibfield
  {journal} {\bibinfo  {journal} {Communications Materials}\ }\textbf {\bibinfo
  {volume} {2}},\ \bibinfo {pages} {18} (\bibinfo {year}
  {2021}{\natexlab{a}})}\BibitemShut {NoStop}%
\bibitem [{\citenamefont {Ikeda}\ \emph
  {et~al.}(2021{\natexlab{b}})\citenamefont {Ikeda}, \citenamefont {Fujiwara},
  \citenamefont {Shiogai}, \citenamefont {Seki}, \citenamefont {Nomura},
  \citenamefont {Takanashi},\ and\ \citenamefont
  {Tsukazaki}}]{Ikeda2021CommPhys}%
  \BibitemOpen
  \bibfield  {author} {\bibinfo {author} {\bibfnamefont {J.}~\bibnamefont
  {Ikeda}}, \bibinfo {author} {\bibfnamefont {K.}~\bibnamefont {Fujiwara}},
  \bibinfo {author} {\bibfnamefont {J.}~\bibnamefont {Shiogai}}, \bibinfo
  {author} {\bibfnamefont {T.}~\bibnamefont {Seki}}, \bibinfo {author}
  {\bibfnamefont {K.}~\bibnamefont {Nomura}}, \bibinfo {author} {\bibfnamefont
  {K.}~\bibnamefont {Takanashi}},\ and\ \bibinfo {author} {\bibfnamefont
  {A.}~\bibnamefont {Tsukazaki}},\ }\bibfield  {title} {\bibinfo {title}
  {Two-dimensionality of metallic surface conduction in co3sn2s2 thin films},\
  }\href {https://doi.org/10.1038/s42005-021-00627-y} {\bibfield  {journal}
  {\bibinfo  {journal} {Communications Physics}\ }\textbf {\bibinfo {volume}
  {4}},\ \bibinfo {pages} {117} (\bibinfo {year}
  {2021}{\natexlab{b}})}\BibitemShut {NoStop}%
\bibitem [{\citenamefont {Kim}\ \emph {et~al.}(2012)\citenamefont {Kim},
  \citenamefont {Sinha}, \citenamefont {Hayashi}, \citenamefont {Yamanouchi},
  \citenamefont {Fukami}, \citenamefont {Suzuki}, \citenamefont {Mitani},\ and\
  \citenamefont {Ohno}}]{KimNatMat2012_Harmonic}%
  \BibitemOpen
  \bibfield  {author} {\bibinfo {author} {\bibfnamefont {J.}~\bibnamefont
  {Kim}}, \bibinfo {author} {\bibfnamefont {J.}~\bibnamefont {Sinha}}, \bibinfo
  {author} {\bibfnamefont {M.}~\bibnamefont {Hayashi}}, \bibinfo {author}
  {\bibfnamefont {M.}~\bibnamefont {Yamanouchi}}, \bibinfo {author}
  {\bibfnamefont {S.}~\bibnamefont {Fukami}}, \bibinfo {author} {\bibfnamefont
  {T.}~\bibnamefont {Suzuki}}, \bibinfo {author} {\bibfnamefont
  {S.}~\bibnamefont {Mitani}},\ and\ \bibinfo {author} {\bibfnamefont
  {H.}~\bibnamefont {Ohno}},\ }\bibfield  {title} {\bibinfo {title} {Layer
  thickness dependence of the current-induced effective field vector in
  {T}a$|${C}o{F}e{B}$|${M}g{O}},\ }\href@noop {} {\bibfield  {journal}
  {\bibinfo  {journal} {Nature Materials}\ }\textbf {\bibinfo {volume} {12}},\
  \bibinfo {pages} {240} (\bibinfo {year} {2012})}\BibitemShut {NoStop}%
\bibitem [{\citenamefont {Garello}\ \emph {et~al.}(2013)\citenamefont
  {Garello}, \citenamefont {Miron}, \citenamefont {Avci}, \citenamefont
  {Freimuth}, \citenamefont {Mokrousov}, \citenamefont {Bl{\"u}gel},
  \citenamefont {Auffret}, \citenamefont {Boulle}, \citenamefont {Gaudin},\
  and\ \citenamefont {Gambardella}}]{GarelloNatNano2013_SOT}%
  \BibitemOpen
  \bibfield  {author} {\bibinfo {author} {\bibfnamefont {K.}~\bibnamefont
  {Garello}}, \bibinfo {author} {\bibfnamefont {I.~M.}\ \bibnamefont {Miron}},
  \bibinfo {author} {\bibfnamefont {C.~O.}\ \bibnamefont {Avci}}, \bibinfo
  {author} {\bibfnamefont {F.}~\bibnamefont {Freimuth}}, \bibinfo {author}
  {\bibfnamefont {Y.}~\bibnamefont {Mokrousov}}, \bibinfo {author}
  {\bibfnamefont {S.}~\bibnamefont {Bl{\"u}gel}}, \bibinfo {author}
  {\bibfnamefont {S.}~\bibnamefont {Auffret}}, \bibinfo {author} {\bibfnamefont
  {O.}~\bibnamefont {Boulle}}, \bibinfo {author} {\bibfnamefont
  {G.}~\bibnamefont {Gaudin}},\ and\ \bibinfo {author} {\bibfnamefont
  {P.}~\bibnamefont {Gambardella}},\ }\bibfield  {title} {\bibinfo {title}
  {Symmetry and magnitude of spin-orbit torques in ferromagnetic
  heterostructures},\ }\href {https://doi.org/10.1038/nnano.2013.145}
  {\bibfield  {journal} {\bibinfo  {journal} {Nature Nanotechnology}\ }\textbf
  {\bibinfo {volume} {8}},\ \bibinfo {pages} {587} (\bibinfo {year}
  {2013})}\BibitemShut {NoStop}%
\bibitem [{\citenamefont {Avci}\ \emph {et~al.}(2014)\citenamefont {Avci},
  \citenamefont {Garello}, \citenamefont {Gabureac}, \citenamefont {Ghosh},
  \citenamefont {Fuhrer}, \citenamefont {Alvarado},\ and\ \citenamefont
  {Gambardella}}]{AvciPRB2014_thermoelectric}%
  \BibitemOpen
  \bibfield  {author} {\bibinfo {author} {\bibfnamefont {C.~O.}\ \bibnamefont
  {Avci}}, \bibinfo {author} {\bibfnamefont {K.}~\bibnamefont {Garello}},
  \bibinfo {author} {\bibfnamefont {M.}~\bibnamefont {Gabureac}}, \bibinfo
  {author} {\bibfnamefont {A.}~\bibnamefont {Ghosh}}, \bibinfo {author}
  {\bibfnamefont {A.}~\bibnamefont {Fuhrer}}, \bibinfo {author} {\bibfnamefont
  {S.~F.}\ \bibnamefont {Alvarado}},\ and\ \bibinfo {author} {\bibfnamefont
  {P.}~\bibnamefont {Gambardella}},\ }\bibfield  {title} {\bibinfo {title}
  {Interplay of spin-orbit torque and thermoelectric effects in
  ferromagnet/normal-metal bilayers},\ }\href@noop {} {\bibfield  {journal}
  {\bibinfo  {journal} {Physical Review B}\ }\textbf {\bibinfo {volume} {90}},\
  \bibinfo {pages} {224427} (\bibinfo {year} {2014})}\BibitemShut {NoStop}%
\bibitem [{\citenamefont {Roschewsky}\ \emph {et~al.}(2019)\citenamefont
  {Roschewsky}, \citenamefont {Walker}, \citenamefont {Gowtham}, \citenamefont
  {Muschinske}, \citenamefont {Hellman}, \citenamefont {Bank},\ and\
  \citenamefont {Salahuddin}}]{RoschewskyPRB2019}%
  \BibitemOpen
  \bibfield  {author} {\bibinfo {author} {\bibfnamefont {N.}~\bibnamefont
  {Roschewsky}}, \bibinfo {author} {\bibfnamefont {E.~S.}\ \bibnamefont
  {Walker}}, \bibinfo {author} {\bibfnamefont {P.}~\bibnamefont {Gowtham}},
  \bibinfo {author} {\bibfnamefont {S.}~\bibnamefont {Muschinske}}, \bibinfo
  {author} {\bibfnamefont {F.}~\bibnamefont {Hellman}}, \bibinfo {author}
  {\bibfnamefont {S.~R.}\ \bibnamefont {Bank}},\ and\ \bibinfo {author}
  {\bibfnamefont {S.}~\bibnamefont {Salahuddin}},\ }\bibfield  {title}
  {\bibinfo {title} {Spin-orbit torque and nernst effect in bi-sb/co
  heterostructures},\ }\href@noop {} {\bibfield  {journal} {\bibinfo  {journal}
  {Phys. Rev. B}\ }\textbf {\bibinfo {volume} {99}},\ \bibinfo {pages} {195103}
  (\bibinfo {year} {2019})}\BibitemShut {NoStop}%
\bibitem [{\citenamefont {Chi}\ \emph {et~al.}(2020)\citenamefont {Chi},
  \citenamefont {Lau}, \citenamefont {Xu}, \citenamefont {Ohkubo},
  \citenamefont {Hono},\ and\ \citenamefont {Hayashi}}]{ChiSciAdv2020_BiSb}%
  \BibitemOpen
  \bibfield  {author} {\bibinfo {author} {\bibfnamefont {Z.}~\bibnamefont
  {Chi}}, \bibinfo {author} {\bibfnamefont {Y.-C.}\ \bibnamefont {Lau}},
  \bibinfo {author} {\bibfnamefont {X.}~\bibnamefont {Xu}}, \bibinfo {author}
  {\bibfnamefont {T.}~\bibnamefont {Ohkubo}}, \bibinfo {author} {\bibfnamefont
  {K.}~\bibnamefont {Hono}},\ and\ \bibinfo {author} {\bibfnamefont
  {M.}~\bibnamefont {Hayashi}},\ }\bibfield  {title} {\bibinfo {title} {The
  spin hall effect of bi-sb alloy driven by thermally excited dirac-like
  electrons},\ }\href@noop {} {\bibfield  {journal} {\bibinfo  {journal}
  {Science Advances}\ }\textbf {\bibinfo {volume} {6}},\ \bibinfo {pages}
  {eaay2324} (\bibinfo {year} {2020})}\BibitemShut {NoStop}%
\bibitem [{\citenamefont {Tang}\ \emph {et~al.}(2021)\citenamefont {Tang},
  \citenamefont {Wen}, \citenamefont {Lau}, \citenamefont {Sukegawa},
  \citenamefont {Seki},\ and\ \citenamefont {Mitani}}]{Tang_APL2021}%
  \BibitemOpen
  \bibfield  {author} {\bibinfo {author} {\bibfnamefont {K.}~\bibnamefont
  {Tang}}, \bibinfo {author} {\bibfnamefont {Z.}~\bibnamefont {Wen}}, \bibinfo
  {author} {\bibfnamefont {Y.-C.}\ \bibnamefont {Lau}}, \bibinfo {author}
  {\bibfnamefont {H.}~\bibnamefont {Sukegawa}}, \bibinfo {author}
  {\bibfnamefont {T.}~\bibnamefont {Seki}},\ and\ \bibinfo {author}
  {\bibfnamefont {S.}~\bibnamefont {Mitani}},\ }\bibfield  {title} {\bibinfo
  {title} {Magnetization switching induced by spin–orbit torque from co2mnga
  magnetic weyl semimetal thin films},\ }\href
  {https://doi.org/10.1063/5.0037178} {\bibfield  {journal} {\bibinfo
  {journal} {Applied Physics Letters}\ }\textbf {\bibinfo {volume} {118}},\
  \bibinfo {pages} {062402} (\bibinfo {year} {2021})}\BibitemShut {NoStop}%
\bibitem [{\citenamefont {Liu}\ \emph {et~al.}(2011)\citenamefont {Liu},
  \citenamefont {Moriyama}, \citenamefont {Ralph},\ and\ \citenamefont
  {Buhrman}}]{LiuPRL2011}%
  \BibitemOpen
  \bibfield  {author} {\bibinfo {author} {\bibfnamefont {L.}~\bibnamefont
  {Liu}}, \bibinfo {author} {\bibfnamefont {T.}~\bibnamefont {Moriyama}},
  \bibinfo {author} {\bibfnamefont {D.~C.}\ \bibnamefont {Ralph}},\ and\
  \bibinfo {author} {\bibfnamefont {R.~A.}\ \bibnamefont {Buhrman}},\
  }\bibfield  {title} {\bibinfo {title} {Spin-torque ferromagnetic resonance
  induced by the spin {H}all effect},\ }\href
  {https://doi.org/10.1103/PhysRevLett.106.036601} {\bibfield  {journal}
  {\bibinfo  {journal} {Physical Review Letters}\ }\textbf {\bibinfo {volume}
  {106}},\ \bibinfo {pages} {036601} (\bibinfo {year} {2011})}\BibitemShut
  {NoStop}%
\bibitem [{\citenamefont {Shen}\ \emph {et~al.}(2020)\citenamefont {Shen},
  \citenamefont {Yao}, \citenamefont {Zeng}, \citenamefont {Sun}, \citenamefont
  {Xi}, \citenamefont {Wu}, \citenamefont {Wang}, \citenamefont {Shen},
  \citenamefont {Liu},\ and\ \citenamefont {Liu}}]{ShenPRL2020}%
  \BibitemOpen
  \bibfield  {author} {\bibinfo {author} {\bibfnamefont {J.}~\bibnamefont
  {Shen}}, \bibinfo {author} {\bibfnamefont {Q.}~\bibnamefont {Yao}}, \bibinfo
  {author} {\bibfnamefont {Q.}~\bibnamefont {Zeng}}, \bibinfo {author}
  {\bibfnamefont {H.}~\bibnamefont {Sun}}, \bibinfo {author} {\bibfnamefont
  {X.}~\bibnamefont {Xi}}, \bibinfo {author} {\bibfnamefont {G.}~\bibnamefont
  {Wu}}, \bibinfo {author} {\bibfnamefont {W.}~\bibnamefont {Wang}}, \bibinfo
  {author} {\bibfnamefont {B.}~\bibnamefont {Shen}}, \bibinfo {author}
  {\bibfnamefont {Q.}~\bibnamefont {Liu}},\ and\ \bibinfo {author}
  {\bibfnamefont {E.}~\bibnamefont {Liu}},\ }\bibfield  {title} {\bibinfo
  {title} {Local disorder-induced elevation of intrinsic anomalous hall
  conductance in an electron-doped magnetic weyl semimetal},\ }\href
  {https://doi.org/10.1103/PhysRevLett.125.086602} {\bibfield  {journal}
  {\bibinfo  {journal} {Phys. Rev. Lett.}\ }\textbf {\bibinfo {volume} {125}},\
  \bibinfo {pages} {086602} (\bibinfo {year} {2020})}\BibitemShut {NoStop}%
\bibitem [{\citenamefont {Belopolski}\ \emph {et~al.}(2019)\citenamefont
  {Belopolski}, \citenamefont {Manna}, \citenamefont {Sanchez}, \citenamefont
  {Chang}, \citenamefont {Ernst}, \citenamefont {Yin}, \citenamefont {Zhang},
  \citenamefont {Cochran}, \citenamefont {Shumiya}, \citenamefont {Zheng},
  \citenamefont {Singh}, \citenamefont {Bian}, \citenamefont {Multer},
  \citenamefont {Litskevich}, \citenamefont {Zhou}, \citenamefont {Huang},
  \citenamefont {Wang}, \citenamefont {Chang}, \citenamefont {Xu},
  \citenamefont {Bansil}, \citenamefont {Felser}, \citenamefont {Lin},\ and\
  \citenamefont {Hasan}}]{Belopolski1278}%
  \BibitemOpen
  \bibfield  {author} {\bibinfo {author} {\bibfnamefont {I.}~\bibnamefont
  {Belopolski}}, \bibinfo {author} {\bibfnamefont {K.}~\bibnamefont {Manna}},
  \bibinfo {author} {\bibfnamefont {D.~S.}\ \bibnamefont {Sanchez}}, \bibinfo
  {author} {\bibfnamefont {G.}~\bibnamefont {Chang}}, \bibinfo {author}
  {\bibfnamefont {B.}~\bibnamefont {Ernst}}, \bibinfo {author} {\bibfnamefont
  {J.}~\bibnamefont {Yin}}, \bibinfo {author} {\bibfnamefont {S.~S.}\
  \bibnamefont {Zhang}}, \bibinfo {author} {\bibfnamefont {T.}~\bibnamefont
  {Cochran}}, \bibinfo {author} {\bibfnamefont {N.}~\bibnamefont {Shumiya}},
  \bibinfo {author} {\bibfnamefont {H.}~\bibnamefont {Zheng}}, \bibinfo
  {author} {\bibfnamefont {B.}~\bibnamefont {Singh}}, \bibinfo {author}
  {\bibfnamefont {G.}~\bibnamefont {Bian}}, \bibinfo {author} {\bibfnamefont
  {D.}~\bibnamefont {Multer}}, \bibinfo {author} {\bibfnamefont
  {M.}~\bibnamefont {Litskevich}}, \bibinfo {author} {\bibfnamefont
  {X.}~\bibnamefont {Zhou}}, \bibinfo {author} {\bibfnamefont {S.-M.}\
  \bibnamefont {Huang}}, \bibinfo {author} {\bibfnamefont {B.}~\bibnamefont
  {Wang}}, \bibinfo {author} {\bibfnamefont {T.-R.}\ \bibnamefont {Chang}},
  \bibinfo {author} {\bibfnamefont {S.-Y.}\ \bibnamefont {Xu}}, \bibinfo
  {author} {\bibfnamefont {A.}~\bibnamefont {Bansil}}, \bibinfo {author}
  {\bibfnamefont {C.}~\bibnamefont {Felser}}, \bibinfo {author} {\bibfnamefont
  {H.}~\bibnamefont {Lin}},\ and\ \bibinfo {author} {\bibfnamefont {M.~Z.}\
  \bibnamefont {Hasan}},\ }\bibfield  {title} {\bibinfo {title} {Discovery of
  topological weyl fermion lines and drumhead surface states in a room
  temperature magnet},\ }\href {https://doi.org/10.1126/science.aav2327}
  {\bibfield  {journal} {\bibinfo  {journal} {Science}\ }\textbf {\bibinfo
  {volume} {365}},\ \bibinfo {pages} {1278} (\bibinfo {year}
  {2019})}\BibitemShut {NoStop}%
\bibitem [{\citenamefont {Manna}\ \emph {et~al.}(2018)\citenamefont {Manna},
  \citenamefont {Muechler}, \citenamefont {Kao}, \citenamefont {Stinshoff},
  \citenamefont {Zhang}, \citenamefont {Gooth}, \citenamefont {Kumar},
  \citenamefont {Kreiner}, \citenamefont {Koepernik}, \citenamefont {Car},
  \citenamefont {K\"ubler}, \citenamefont {Fecher}, \citenamefont {Shekhar},
  \citenamefont {Sun},\ and\ \citenamefont {Felser}}]{MannaPRX2018}%
  \BibitemOpen
  \bibfield  {author} {\bibinfo {author} {\bibfnamefont {K.}~\bibnamefont
  {Manna}}, \bibinfo {author} {\bibfnamefont {L.}~\bibnamefont {Muechler}},
  \bibinfo {author} {\bibfnamefont {T.-H.}\ \bibnamefont {Kao}}, \bibinfo
  {author} {\bibfnamefont {R.}~\bibnamefont {Stinshoff}}, \bibinfo {author}
  {\bibfnamefont {Y.}~\bibnamefont {Zhang}}, \bibinfo {author} {\bibfnamefont
  {J.}~\bibnamefont {Gooth}}, \bibinfo {author} {\bibfnamefont
  {N.}~\bibnamefont {Kumar}}, \bibinfo {author} {\bibfnamefont
  {G.}~\bibnamefont {Kreiner}}, \bibinfo {author} {\bibfnamefont
  {K.}~\bibnamefont {Koepernik}}, \bibinfo {author} {\bibfnamefont
  {R.}~\bibnamefont {Car}}, \bibinfo {author} {\bibfnamefont {J.}~\bibnamefont
  {K\"ubler}}, \bibinfo {author} {\bibfnamefont {G.~H.}\ \bibnamefont
  {Fecher}}, \bibinfo {author} {\bibfnamefont {C.}~\bibnamefont {Shekhar}},
  \bibinfo {author} {\bibfnamefont {Y.}~\bibnamefont {Sun}},\ and\ \bibinfo
  {author} {\bibfnamefont {C.}~\bibnamefont {Felser}},\ }\bibfield  {title}
  {\bibinfo {title} {From colossal to zero: Controlling the anomalous hall
  effect in magnetic heusler compounds via berry curvature design},\ }\href
  {https://doi.org/10.1103/PhysRevX.8.041045} {\bibfield  {journal} {\bibinfo
  {journal} {Phys. Rev. X}\ }\textbf {\bibinfo {volume} {8}},\ \bibinfo {pages}
  {041045} (\bibinfo {year} {2018})}\BibitemShut {NoStop}%
\bibitem [{\citenamefont {Markou}\ \emph {et~al.}(2021)\citenamefont {Markou},
  \citenamefont {Gayles}, \citenamefont {Derunova}, \citenamefont {Swekis},
  \citenamefont {Noky}, \citenamefont {Zhang}, \citenamefont {Ali},
  \citenamefont {Sun},\ and\ \citenamefont {Felser}}]{Markou2021}%
  \BibitemOpen
  \bibfield  {author} {\bibinfo {author} {\bibfnamefont {A.}~\bibnamefont
  {Markou}}, \bibinfo {author} {\bibfnamefont {J.}~\bibnamefont {Gayles}},
  \bibinfo {author} {\bibfnamefont {E.}~\bibnamefont {Derunova}}, \bibinfo
  {author} {\bibfnamefont {P.}~\bibnamefont {Swekis}}, \bibinfo {author}
  {\bibfnamefont {J.}~\bibnamefont {Noky}}, \bibinfo {author} {\bibfnamefont
  {L.}~\bibnamefont {Zhang}}, \bibinfo {author} {\bibfnamefont {M.~N.}\
  \bibnamefont {Ali}}, \bibinfo {author} {\bibfnamefont {Y.}~\bibnamefont
  {Sun}},\ and\ \bibinfo {author} {\bibfnamefont {C.}~\bibnamefont {Felser}},\
  }\bibfield  {title} {\bibinfo {title} {Hard magnet topological semimetals in
  xpt3 compounds with the harmony of berry curvature},\ }\href
  {https://doi.org/10.1038/s42005-021-00608-1} {\bibfield  {journal} {\bibinfo
  {journal} {Communications Physics}\ }\textbf {\bibinfo {volume} {4}},\
  \bibinfo {pages} {104} (\bibinfo {year} {2021})}\BibitemShut {NoStop}%
\bibitem [{\citenamefont {Sakai}\ \emph {et~al.}(2020)\citenamefont {Sakai},
  \citenamefont {Minami}, \citenamefont {Koretsune}, \citenamefont {Chen},
  \citenamefont {Higo}, \citenamefont {Wang}, \citenamefont {Nomoto},
  \citenamefont {Hirayama}, \citenamefont {Miwa}, \citenamefont
  {Nishio-Hamane}, \citenamefont {Ishii}, \citenamefont {Arita},\ and\
  \citenamefont {Nakatsuji}}]{Sakai2020}%
  \BibitemOpen
  \bibfield  {author} {\bibinfo {author} {\bibfnamefont {A.}~\bibnamefont
  {Sakai}}, \bibinfo {author} {\bibfnamefont {S.}~\bibnamefont {Minami}},
  \bibinfo {author} {\bibfnamefont {T.}~\bibnamefont {Koretsune}}, \bibinfo
  {author} {\bibfnamefont {T.}~\bibnamefont {Chen}}, \bibinfo {author}
  {\bibfnamefont {T.}~\bibnamefont {Higo}}, \bibinfo {author} {\bibfnamefont
  {Y.}~\bibnamefont {Wang}}, \bibinfo {author} {\bibfnamefont {T.}~\bibnamefont
  {Nomoto}}, \bibinfo {author} {\bibfnamefont {M.}~\bibnamefont {Hirayama}},
  \bibinfo {author} {\bibfnamefont {S.}~\bibnamefont {Miwa}}, \bibinfo {author}
  {\bibfnamefont {D.}~\bibnamefont {Nishio-Hamane}}, \bibinfo {author}
  {\bibfnamefont {F.}~\bibnamefont {Ishii}}, \bibinfo {author} {\bibfnamefont
  {R.}~\bibnamefont {Arita}},\ and\ \bibinfo {author} {\bibfnamefont
  {S.}~\bibnamefont {Nakatsuji}},\ }\bibfield  {title} {\bibinfo {title}
  {Iron-based binary ferromagnets for transverse thermoelectric conversion},\
  }\href {https://doi.org/10.1038/s41586-020-2230-z} {\bibfield  {journal}
  {\bibinfo  {journal} {Nature}\ }\textbf {\bibinfo {volume} {581}},\ \bibinfo
  {pages} {53} (\bibinfo {year} {2020})}\BibitemShut {NoStop}%
\bibitem [{\citenamefont {Asaba}\ \emph {et~al.}(2021)\citenamefont {Asaba},
  \citenamefont {Ivanov}, \citenamefont {Thomas}, \citenamefont {Savrasov},
  \citenamefont {Thompson}, \citenamefont {Bauer},\ and\ \citenamefont
  {Ronning}}]{Asaba_SA2021}%
  \BibitemOpen
  \bibfield  {author} {\bibinfo {author} {\bibfnamefont {T.}~\bibnamefont
  {Asaba}}, \bibinfo {author} {\bibfnamefont {V.}~\bibnamefont {Ivanov}},
  \bibinfo {author} {\bibfnamefont {S.~M.}\ \bibnamefont {Thomas}}, \bibinfo
  {author} {\bibfnamefont {S.~Y.}\ \bibnamefont {Savrasov}}, \bibinfo {author}
  {\bibfnamefont {J.~D.}\ \bibnamefont {Thompson}}, \bibinfo {author}
  {\bibfnamefont {E.~D.}\ \bibnamefont {Bauer}},\ and\ \bibinfo {author}
  {\bibfnamefont {F.}~\bibnamefont {Ronning}},\ }\bibfield  {title} {\bibinfo
  {title} {Colossal anomalous nernst effect in a correlated noncentrosymmetric
  kagome ferromagnet},\ }\href {https://doi.org/10.1126/sciadv.abf1467}
  {\bibfield  {journal} {\bibinfo  {journal} {Science Advances}\ }\textbf
  {\bibinfo {volume} {7}},\ \bibinfo {pages} {eabf1467} (\bibinfo {year}
  {2021})}\BibitemShut {NoStop}%
\bibitem [{\citenamefont {Pan}\ \emph {et~al.}(2022)\citenamefont {Pan},
  \citenamefont {Le}, \citenamefont {He}, \citenamefont {Watzman},
  \citenamefont {Yao}, \citenamefont {Gooth}, \citenamefont {Heremans},
  \citenamefont {Sun},\ and\ \citenamefont {Felser}}]{Pan2022}%
  \BibitemOpen
  \bibfield  {author} {\bibinfo {author} {\bibfnamefont {Y.}~\bibnamefont
  {Pan}}, \bibinfo {author} {\bibfnamefont {C.}~\bibnamefont {Le}}, \bibinfo
  {author} {\bibfnamefont {B.}~\bibnamefont {He}}, \bibinfo {author}
  {\bibfnamefont {S.~J.}\ \bibnamefont {Watzman}}, \bibinfo {author}
  {\bibfnamefont {M.}~\bibnamefont {Yao}}, \bibinfo {author} {\bibfnamefont
  {J.}~\bibnamefont {Gooth}}, \bibinfo {author} {\bibfnamefont {J.~P.}\
  \bibnamefont {Heremans}}, \bibinfo {author} {\bibfnamefont {Y.}~\bibnamefont
  {Sun}},\ and\ \bibinfo {author} {\bibfnamefont {C.}~\bibnamefont {Felser}},\
  }\bibfield  {title} {\bibinfo {title} {Giant anomalous nernst signal in the
  antiferromagnet ybmnbi2},\ }\href
  {https://doi.org/10.1038/s41563-021-01149-2} {\bibfield  {journal} {\bibinfo
  {journal} {Nature Materials}\ }\textbf {\bibinfo {volume} {21}},\ \bibinfo
  {pages} {203} (\bibinfo {year} {2022})}\BibitemShut {NoStop}%
\bibitem [{\citenamefont {Meyer}\ \emph {et~al.}(2017)\citenamefont {Meyer},
  \citenamefont {Chen}, \citenamefont {Wimmer}, \citenamefont {Althammer},
  \citenamefont {Wimmer}, \citenamefont {Schlitz}, \citenamefont {Gepr{\"a}gs},
  \citenamefont {Huebl}, \citenamefont {K{\"o}dderitzsch}, \citenamefont
  {Ebert}, \citenamefont {Bauer}, \citenamefont {Gross},\ and\ \citenamefont
  {Goennenwein}}]{Meyer2017}%
  \BibitemOpen
  \bibfield  {author} {\bibinfo {author} {\bibfnamefont {S.}~\bibnamefont
  {Meyer}}, \bibinfo {author} {\bibfnamefont {Y.-T.}\ \bibnamefont {Chen}},
  \bibinfo {author} {\bibfnamefont {S.}~\bibnamefont {Wimmer}}, \bibinfo
  {author} {\bibfnamefont {M.}~\bibnamefont {Althammer}}, \bibinfo {author}
  {\bibfnamefont {T.}~\bibnamefont {Wimmer}}, \bibinfo {author} {\bibfnamefont
  {R.}~\bibnamefont {Schlitz}}, \bibinfo {author} {\bibfnamefont
  {S.}~\bibnamefont {Gepr{\"a}gs}}, \bibinfo {author} {\bibfnamefont
  {H.}~\bibnamefont {Huebl}}, \bibinfo {author} {\bibfnamefont
  {D.}~\bibnamefont {K{\"o}dderitzsch}}, \bibinfo {author} {\bibfnamefont
  {H.}~\bibnamefont {Ebert}}, \bibinfo {author} {\bibfnamefont {G.~E.~W.}\
  \bibnamefont {Bauer}}, \bibinfo {author} {\bibfnamefont {R.}~\bibnamefont
  {Gross}},\ and\ \bibinfo {author} {\bibfnamefont {S.~T.~B.}\ \bibnamefont
  {Goennenwein}},\ }\bibfield  {title} {\bibinfo {title} {Observation of the
  spin nernst effect},\ }\href {https://doi.org/10.1038/nmat4964} {\bibfield
  {journal} {\bibinfo  {journal} {Nature Materials}\ }\textbf {\bibinfo
  {volume} {16}},\ \bibinfo {pages} {977} (\bibinfo {year} {2017})}\BibitemShut
  {NoStop}%
\bibitem [{\citenamefont {Sheng}\ \emph {et~al.}(2017)\citenamefont {Sheng},
  \citenamefont {Sakuraba}, \citenamefont {Lau}, \citenamefont {Takahashi},
  \citenamefont {Mitani},\ and\ \citenamefont {Hayashi}}]{SNE_Sheng}%
  \BibitemOpen
  \bibfield  {author} {\bibinfo {author} {\bibfnamefont {P.}~\bibnamefont
  {Sheng}}, \bibinfo {author} {\bibfnamefont {Y.}~\bibnamefont {Sakuraba}},
  \bibinfo {author} {\bibfnamefont {Y.-C.}\ \bibnamefont {Lau}}, \bibinfo
  {author} {\bibfnamefont {S.}~\bibnamefont {Takahashi}}, \bibinfo {author}
  {\bibfnamefont {S.}~\bibnamefont {Mitani}},\ and\ \bibinfo {author}
  {\bibfnamefont {M.}~\bibnamefont {Hayashi}},\ }\bibfield  {title} {\bibinfo
  {title} {The spin nernst effect in tungsten},\ }\href
  {https://doi.org/10.1126/sciadv.1701503} {\bibfield  {journal} {\bibinfo
  {journal} {Science Advances}\ }\textbf {\bibinfo {volume} {3}},\ \bibinfo
  {pages} {e1701503} (\bibinfo {year} {2017})}\BibitemShut {NoStop}%
\bibitem [{\citenamefont {Pu}\ \emph {et~al.}(2008)\citenamefont {Pu},
  \citenamefont {Chiba}, \citenamefont {Matsukura}, \citenamefont {Ohno},\ and\
  \citenamefont {Shi}}]{PuPRL2008}%
  \BibitemOpen
  \bibfield  {author} {\bibinfo {author} {\bibfnamefont {Y.}~\bibnamefont
  {Pu}}, \bibinfo {author} {\bibfnamefont {D.}~\bibnamefont {Chiba}}, \bibinfo
  {author} {\bibfnamefont {F.}~\bibnamefont {Matsukura}}, \bibinfo {author}
  {\bibfnamefont {H.}~\bibnamefont {Ohno}},\ and\ \bibinfo {author}
  {\bibfnamefont {J.}~\bibnamefont {Shi}},\ }\bibfield  {title} {\bibinfo
  {title} {Mott relation for anomalous hall and nernst effects in
  ${\mathrm{ga}}_{1\ensuremath{-}x}{\mathrm{mn}}_{x}\mathrm{As}$ ferromagnetic
  semiconductors},\ }\href {https://doi.org/10.1103/PhysRevLett.101.117208}
  {\bibfield  {journal} {\bibinfo  {journal} {Phys. Rev. Lett.}\ }\textbf
  {\bibinfo {volume} {101}},\ \bibinfo {pages} {117208} (\bibinfo {year}
  {2008})}\BibitemShut {NoStop}%
\end{thebibliography}


\end{document}